\documentclass[12pt]{article}
\usepackage{graphicx}
\usepackage{color}
\usepackage{times}
\usepackage{amsmath}
\usepackage{amsfonts}
\usepackage{enumerate}
\usepackage{lscape}
\usepackage{booktabs}
\usepackage{epstopdf}
\usepackage[margin=0.75in]{geometry}
\usepackage{hyperref}
\usepackage[normalem]{ulem}
\newcommand{\et}{\textit{et. al}}

\def\boxit#1{\vbox{\hrule\hbox{\vrule\kern6pt
          \vbox{\kern6pt#1\kern6pt}\kern6pt\vrule}\hrule}}

\def\boxit#1{\vbox{\hrule\hbox{\vrule\kern6pt
          \vbox{\kern6pt#1\kern6pt}\kern6pt\vrule}\hrule}}

\def\bse{\begin{eqnarray*}}
\def\ese{\end{eqnarray*}}
\def\be{\begin{eqnarray}}
\def\ee{\end{eqnarray}}
\def\bq{\begin{equation}}
\def\eq{\end{equation}}
\def\bse{\begin{eqnarray*}}
\def\ese{\end{eqnarray*}}

\def\boxit#1{\vbox{\hrule\hbox{\vrule\kern6pt
          \vbox{\kern6pt#1\kern6pt}\kern6pt\vrule}\hrule}}

\def\bse{\begin{eqnarray*}}
\def\ese{\end{eqnarray*}}
\def\be{\begin{eqnarray}}
\def\ee{\end{eqnarray}}
\def\bq{\begin{equation}}
\def\eq{\end{equation}}
\def\bse{\begin{eqnarray*}}
\def\ese{\end{eqnarray*}}

\def\boldphi{\mbox{\boldmath $\phi$}}

\def\bphi{\boldphi}
\def\bpsi{\mbox{\boldmath $\psi$}}
\def\bt{\mbox{\boldmath $t$}}

\def\boldU{{\bf U}}
\def\bU{{\boldU}}

\def\boldPhi{\mbox{\boldmath $\Phi$}}

\def\bPhi{\boldPhi}
\def\bPsi{\mbox{\boldmath $\Psi$}}

\def\bB{{\bf B}}

\def\path{\mbox{p}}

\definecolor{DarkGreen}{rgb}{0,0.5,0.15}

\def\squarebox#1{\hbox to #1{\hfill\vbox to #1{\vfill}}}
\def\boxit#1{\vbox{\hrule\hbox{\vrule\kern6pt
          \vbox{\kern6pt#1\kern6pt}\kern6pt\vrule}\hrule}}

\def\be{\mbox{Be}}

\def\bX{\mbox{\bf X}}

\begin{document}
\setcounter{page}{0}
%\begin{titlepage}

\title{Functional Regression}
\author{Jeffrey S. Morris}

\maketitle 
\begin{center}
The University of Texas, MD Anderson Cancer Center

Unit 1411

PO Box 301402

Houston, TX 77230-1402

jefmorris@mdanderson.org
\end{center}

%\end{titlepage}
\begin{abstract}
Functional data analysis (FDA) involves the analysis of data whose ideal units of observation are functions defined on some continuous domain, and the observed data consist of a sample of functions taken from some population, sampled on a discrete grid.  Ramsay and Silverman's 1997 textbook sparked the development of this field, which has accelerated in the past 10 years to become one of the fastest growing areas of statistics, fueled by the growing number of applications yielding this type of data.   One unique characteristic of FDA is the need to combine information both across and within functions, which Ramsay and Silverman called replication and regularization, respectively.  This article will focus on functional regression, the area of FDA that has received the most attention in applications and methodological development.  First will be an introduction to basis functions, key building blocks for regularization in functional regression methods, followed by an overview of functional regression methods, split into three types: [1] functional predictor regression (scalar-on-function), [2] functional response regression (function-on-scalar) and [3] function-on-function regression.  For each, the role of replication and regularization will be discussed and the methodological development described in a roughly chronological manner, at times deviating from the historical timeline to group together similar methods.  The primary focus is on modeling and methodology, highlighting the modeling structures that have been developed and the various regularization approaches employed.  At the end is a brief discussion describing potential areas of future development in this field.

\vspace{2in}

\end{abstract}

\thispagestyle{empty}

\newpage

\section{Introduction}

In recent decades, technological innovations have produced data that are increasingly complex, high dimensional, and structured.   %There is a dire need for new statistical tools for analyzing these data, which require flexibility to capture their rich structure yet computationally efficiency to scale up to their sometimes enormous size.  
A large fraction of these data can be characterized as \textit{functional data}, and has sparked the rapid development of a new area of statistics, \textit{functional data analysis} (FDA).

With functional data, the ideal units of observation are functions defined on some continuous domain, and the observed data consist of a sample of functions taken from some population, with each function sampled on a discrete grid.  This grid could be fine or sparse, regular or irregular, common or varying across sampled functions.  Most commonly, these functions are defined on a 1-dimensional Euclidean domain, but this paradigm also includes functions on higher dimensional domains such as 2d or 3d image data or time-space data, as well as functions observed on manifolds and other non-Euclidean domains.    The genesis of this field was largely motivated by {\em longitudinal data}, growth curves or time series for which $t$ represents time, typically observed on sparse grids.  Recently, technological innovations in various application areas  have produced functions on other domains, including spatial domains, imaging domains, spectral domains, and even genomic locations, which has brought with it higher dimensional and more complex functional data.
The key idea characterizing FDA is to think of each function as a single structured object rather than a collection of data points, which engenders a simplicity of thought that enables the building of models that can simultaneously handle complex structure both within functions and between functions.     %%% Applications include, ..., cite RS applications textbook.
 %partition the modeling problem into its within-function and between-function components, and thus deal with the considerable complexity that can be encountered with these data.

%Many commonly used analysis approaches for these data can be thought of as naive and fail to use all of the information in the data.  Two commonly used approaches at the extremes are \textit{element-wise modeling}, in which different measurements within the functions are modeled independently, and \textit{feature extraction}, wherein simple summaries are computed from each functions and only those analyzed.  They both have significant weaknesses.  The feature extraction approach throws away any information not contained in the features, and element-wise modeling models all the data ignores the relationships between observations within the function, which leads to inefficient and sometimes incorrect inference.  New methods developed for FDA attempt to find a middle ground, modeling the functions as they are but using statistical modeling tools to borrow strength between observations within a function to gain efficiency and stability in estimation and inference.

The recognition of FDA as its own field was sparked in large part by the seminal textbook of Ramsay \& Silverman (1997).  This book introduced the philosophical thinking behind FDA, and described many aspects of FDA including registration, clustering, graphical displays, descriptive statistics, and various strategies for regression analysis.  In their final chapter, they made a key observation that two factors are simultaneously at work in FDA: \textit{replication} and \textit{regularization}.  Replication involves combining information \textit{across} functions to draw upon their commonalities and make inferences on the populations from which they were sampled, while regularization involves the borrowing of strength \textit{within} a function, exploiting the expected underlying structural relationships within a function to gain efficiency and interpretability.  

This article will focus on functional regression, the area of FDA that has gotten the most attention in applications and methodological development.  The goal of this article is to describe the historical development of work in this area, and to highlight the role of regularization and replication in the various modeling approaches that have been developed.  First will be an overview of basis functions, key building blocks for functional regression methods, followed by a description of the general strategies used for regularization in functional regression using various basis functions.  Next, the methodological developments for each type of functional regression will be described in turn: [1] functional predictor regression (scalar-on-function), [2] functional response regression (function-on-scalar), and [3] function-on-function regression.  Each section will begin with a general discussion of the role of replication and regularization in the given type of regression, and then step through the methodological development in a historical manner, at times deviating from the historical timeline to mention closely related work or describe a cluster of work done by the same researchers.  The primary focus will be on modeling and methodology, not so much on theory or computations, highlighting the modeling structures that have been developed and the various regularization approaches employed.  The end of each section will contain a summary and a list of publicly available software for fitting some of the methods discussed.  

While much work has focused on point estimation, some papers have highlighted the inferential capabilities of their methods, and in these cases inferential approaches will be discussed, which include point wise and joint intervals, variable selection, and hypothesis testing, based on either asymptotic results, bootstrap, or Bayesian modeling.  Most methods involve linear regression and  independently sampled 1d Euclidean functions, but others are more advanced, handling nonlinear regression, correlated functions, and functions on higher dimensional domains like images. The presentation will highlight these advances.

\section{Example Data Sets} \label{sec:data}

\begin{figure}
\centering
\centerline{\includegraphics[height=5in]{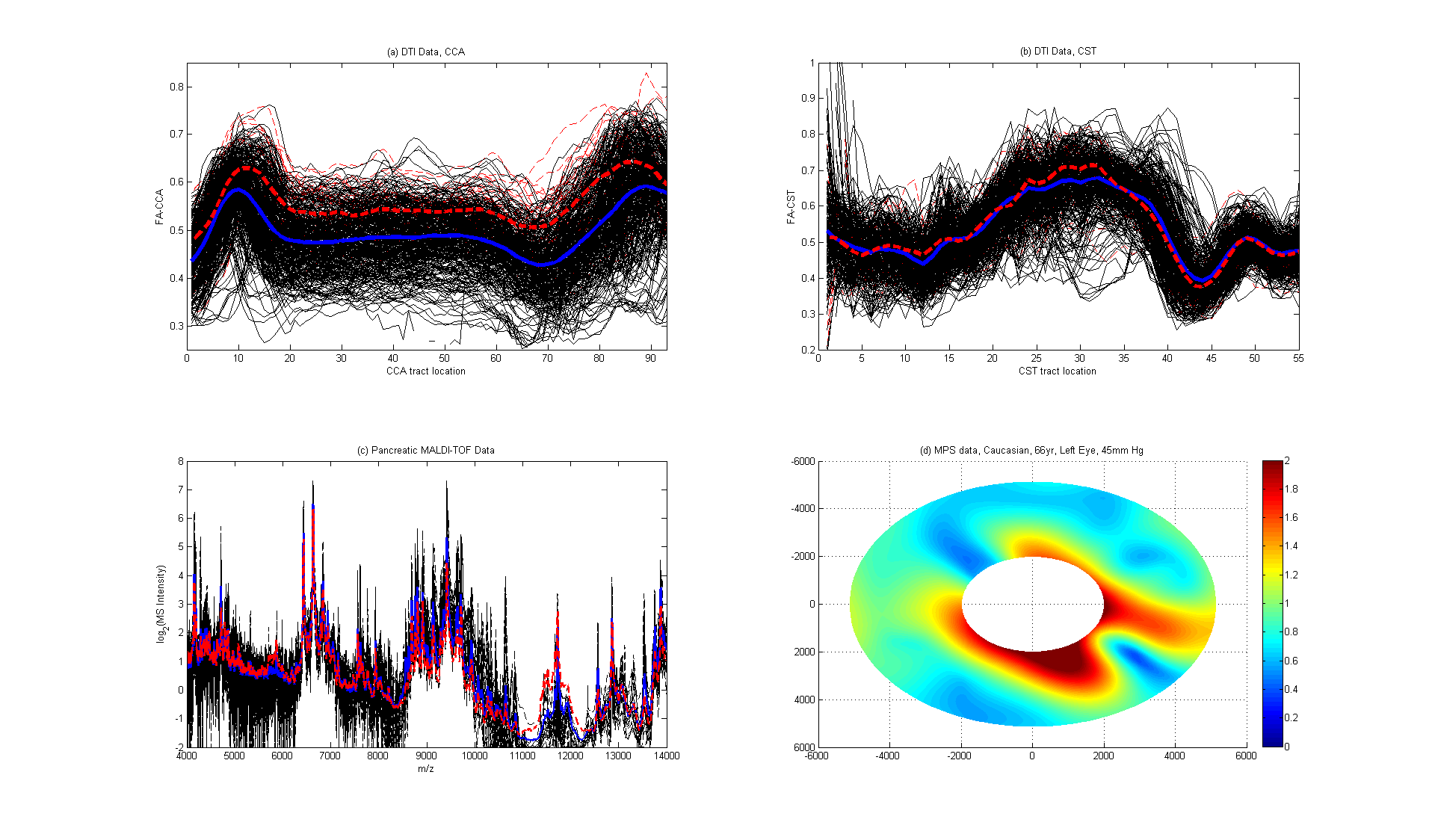}}
\caption{\small{\textbf{Example data sets.}  (a) Fractional anisotropy (FA) measurements from corpus collusum (CCA) and (b) right corticospinal tract (RCST) from the \textit{DTI data set}. Curves from MS patients are plotted in black, and from control patients in red, with mean MS curve in blue and mean control curve in red.  (c) Log spectral intensities from the \textit{mass spectrometry data set}, plotting part of spectrum from 4,000D to 14,000D.  Black lines are plotted spectra from 20 pancreatic cancer (solid) and 20 control (dashed) patients, with mean spectra for pancreatic cancer (red) and control (blue). (d) polar azimuthal projection of partial spherical MPS function for the left eye from one subject, age 66yr and Caucasian, in the \textit{glaucoma data set}, under 45mmHg of interocular pressure.} }\label{fig:data}
\end{figure}

In this section, three functional data sets are introduced to highlight some different types of functional data and provide a concrete backdrop for the discussion of functional regression methods.  For the first two, functions are defined on a 1d Euclidean domain, with the first (\textit{DTI data set}) having relatively simple, smooth functions sampled on a moderately sized grid ($\approx 50-100$ observations per curve), and the second (\textit{mass spectrometry data set}) having complex, spiky functions sampled on a high dimensional grid ($\approx 12,000$ observations per curve).  The third (\textit{glaucoma data set}) is included to illustrate functional data on non-Euclidean domains, with functions defined on a 2d partial spherical domain.  Figure \ref{fig:data} contains representative plots of these data sets.
%, with (a) and (b) plotting all of the curves from the DTI data set, (c) plotting 40 out of 256 curves for the mass spectrometry data set, and (d) plotting one of the 342 functions for the glaucoma data.  
One can refer back to these example data in thinking about the different functional regression methods and associated modeling issues described in this article.%, and they illustrate some of the breadth of data types falling within the FDA paradigm.

\textbf{\underline{DTI Data Set}} (simple, smooth functions, moderate grid):  This data set contains fractional anisotropy (FA) tract profiles from diffusion tensor imaging (DTI) data collected at Johns Hopkins University and the Kennedy-Krieger Institute, and is freely available as part of the $refund$ R package.  The data set involves 382 DTI scans from a total of 142 subjects, 100 with multiple sclerosis (MS) and 42 healthy controls.  From each DTI scan, FA profiles are obtained along the corpus callosum (FA-CCA) and right corticospinal (FA-RCST) tracts, yielding functions $Y^{CCA}_i(t)$ and $Y^{RCST}_i(t), i=1, \ldots, N=382$.  Additionally, for each subject the data includes their MS status $X^{MS}_i=1$ if MS, $0$ if control, plus their Paced Auditory Serial Additional Test (pasat) score $X^{PASAT}_i$, a continuous measurement of cognitive function relating to speed and flexibility in processing auditory information.    As seen in Figure \ref{fig:data}(a)-(b), these functions are relatively simple and smooth and sampled on a moderately sized  grid of 93 (for CCA) or 55 (for RCST) positions.

This data set can be used to illustrate all three types of functional regression described in this article, and has been used in numerous papers in current literature (Greven et al. 2010, Goldsmith et al. 2011, Staicu et al. 2011, Ivanescu et al. 2012, and Scheipl et al. 2014).  To predict MS status from the FA-CCA, one could perform \textbf{functional predictor regression} (Section \ref{sec:FPR}) by regressing MS status $X^{MS}_i$ on $Y^{CCA}_i(t)$ in equation (\ref{eq:GFLM}).   Using \textbf{functional response regression} (Section \ref{sec:FRR}), one could assess differences in the mean FA-CCA profiles between MS and control patients by regressing $Y^{CCA}_i(t)$ on $X^{MS}_i$ in equation (\ref{eq:FRR1})  with functional coefficient $B(t)$ representing the difference in mean FA-CCA profile between MS and control subjects, and then either testing whether $B(t)=0$ or using point wise or joint intervals to characterize for which regions of $t$ the groups differ.    If multiple profiles per subject are modeled, then \textit{random effect functions} should be added to the model to form a functional mixed model like equation (\ref{eq:FMM1}) to account for the correlation between curves from the same subject.  After regressing $Y^{CCA}_i(t)$ on $X^{MS}_i$, one could predict future subjects' MS status from their FA-CCA profiles using \textbf{functional discriminant analysis} (Section \ref{sec:FRR}) as an alternative to functional predictor regression.  One could also regress $Y^{CCA}_i(t)$ on $X^{PASAT}_i$ to estimate a functional slope $B^{PASAT}(t)$ that measures the linear association between FA-CCA at position $t$ and pasat score, or using semiparametric functional mixed models (Section \ref{sec:FRR}) one could nonparametrically model a surface $f^{PASAT}(x, t)$ that measures the smooth effect of pasat ($x$) on FA-CCA at position $t$ in the curve.  To relate the FA-CCA and FA-RCST curves to each other, one could use \textbf{function-on-function regression} (Section \ref{sec:FFR}) to regress $Y^{CCA}_i(t)$ on $Y^{RCST}_i(s)$ using equation (\ref{eq:FonF}) to estimate and perform inference on the resulting regression surface $B(s,t)$.

\textbf{ \underline{ Mass Spectrometry Data Set}} (complex, spiky high-dimensional functions): This data set contains mass spectra from a study to find serum proteomic markers for pancreatic cancer performed at the University of Texas MD Anderson Cancer Center.  As described in Koomen et al. (2005), blood serum was taken from 139 pancreatic cancer patients and 117 healthy controls, and run on a mass spectrometry instrument to obtain proteomic spectra $Y_i(t)$ for each sample $i=1,\ldots,256$.  Each spectrum is a spiky function whose peaks correspond to proteins and peptides with molecular mass (actually mass per unit charge m/z) roughly equal to $t$, and whose spectral intensity approximates its relative abundance.  Each spectrum is observed on a high dimensional grid of size $T=12,096$ for a range of m/z values from $4,000D$ to $40,000D$.  As can be see in in Figure \ref{fig:data}(c), these functions are spiky and irregular, and illustrate a type of complex, high-dimensional functional data that is not the focus of most functional regression development, but increasingly encountered in many applications.  For data like these, flexibility, adaptiveness, and computational efficiency are important considerations for methodology development.

This data set can be used to illustrate both functional predictor regression and functional response regression, and has been used in Morris, et al. (2008), Zhu et al. (2011), Zhu et al. (2012), and Morris (2012).   Frequently, mass spectrometry data are analyzed by performing peak detection and quantification and then surveying the peaks, but this approach discards information in the spectra as no peak detection algorithm is perfect.  Morris, et al. (2008) and Morris (2012) demonstrated that modeling the entire spectrum as a function can yield more power for detecting differentially expressed proteins than peak detection based approaches.   \textbf{Functional response regression} can be performed to regress $Y_i(t)$ on $X^{cancer}_i(=1$ if from cancer patient, $=0$ if control) using equation (\ref{eq:FRR1}) and then assessing for which $t$ is $B(t)$ different from 0, where $B(t)$ represents the difference between cancer and control mean spectra.  After fitting this model, \textbf{functional discriminant analysis} can be done to classify future subjects based on spectra from their serum samples.  Alternatively, one could predict cancer status using \textbf{functional predictor regression} to regress cancer status $X^{cancer}_i$ on spectra $Y_i(t)$ using one of the methods based on equation (\ref{eq:GFLM}).  

\textbf{\underline{Glaucoma Data Set}} (functions on a partial spherical manifold): This data set is from a glaucoma study investigating biomechanics of the scleral surface of the eye (Fazio et al. 2013).  Using custom instrumentation,  investigators were able to induce a fixed amount of interocular pressure and measure scleral displacement continuously around the eye.  Their hypothesis is that individuals at higher risk of glaucoma, including older subjects and African Americans,  should have lower displacement, especially closer to the optic nerve.  To study this, for a series of 9 interocular pressures (7mm, 10mm, 15mm, 20mm, 25mm, 30mm, 35mm, 40mm, and 45mm Hg), the displacement for left and right eyes for 19 donors was measured yielding functions $Y_i(\theta, \phi), i=1,\ldots,N=19 \times 2 \times 9=342$ representing the maximum principal strain (MPS) at 120 circumferential locations $\phi \in (0^{\circ}, 360^{\circ})$ and 120 meridional locations $\theta \in (9^{\circ}, 24^{\circ})$, where $\theta=0^{\circ}$ corresponds to the optic nerve.  These 2d functions are defined on a fine grid (size 14,400) on a partial spherical manifold.  Figure \ref{fig:data}(d) plots a polar azimuthal projection of one of the functions.  For each eye, the subject's age $X^{age}_i$ and race $X^{AA}_i=1$ if African American, $0$ otherwise, is also obtained.

\textbf{Functional response regression} of $Y_i(\theta, \phi)$ on $X^{AA}_i$ yields functional coefficient $B^{AA}(\theta, \phi)$ that can be investigated to test whether MPS differs by race and, if so, in what regions of the eye, and including $X^{age}_i$ yields a functional slope coefficient $B^{age}(\theta, \phi)$ to assess how MPS changes with aging.  Functional mixed models (Section \ref{sec:FRR}) need to be used to account for correlation between eyes from the same subject, and for longitudinal correlations across measurements for different pressure levels in the same eye.  If not wanting to assume linearity in age, semiparametric functional mixed models can be used to estimate $f(age, \theta, \phi)$ that captures the effect of age on MPS as a smooth nonparametric effect varying across locations within the eye, indexed by $(\theta, \phi)$.  One such modeling effort for these data is presented in Lee et al. (2014).  This example illustrates how functional regression approaches can simultaneously handle complex within- and across-function structure and be applied to functions on non-Euclidean domains.

\section{Replication, Regularization, and Basis Functions} \label{sec:basis}

Inherently, any statistical method combines information across observations in some way to make inference on the populations from which they were sampled.  Most basic methods involve either \textit{replication}, combining information \textit{across} sampled units, or \textit{regularization}, combining information \textit{within} sampled units.  FDA involves both.

To expound, replication is the concept exploited in most commonly used statistical models.  For example, given replicate sampled pairs of responses and predictors, regression models combine information across these pairs to discover the relationship between them and build a predictive model.   Replication is also the underlying concept involved when accounting for correlation among the sampled observations, e.g. induced by multi-level designs, in which case information is combined together in a careful way depending on the relative variabilities and sample sizes at each stochastic level, for example using mixed effect models or Bayesian hierarchical models.  In FDA, we consider each function as the sampled unit, and replication involves how our models combine information across functions so we can make inferences on the populations from which they were drawn.

Regularization involves borrowing strength across observations within a function that are expected to be similar to each other, exploiting their expected regularity even though they are discretely observed.  This is the key idea behind (single function) nonparametric regression or kriging in spatial data analysis, where estimation of a process at some location $t$ is improved by borrowing strength from nearby observations within the function using some smoothing mechanism.   For most FDA methods, regularization involves global smoothing, implying an assumption that similarity within the function is strictly related to distance between the observations.  Global smoothing may not be sufficient to adequately capture the internal structure of some functions, especially more complex high dimensional ones such as the \textit{mass spectrometry data} in Section \ref{sec:data}.  For example, some functions are \textit{spatially heterogeneous}, meaning some parts of the function are smoother (i.e. more similar across $t$) than others, and in other cases there are distant parts of the function that are correlated with each other.  In these cases, more general forms of regularization are needed.  Regularization is closely tied to the modeling of correlation for Gaussian data.  It can be applied in the likelihood of the functional data to capture the within-function correlation, or in penalties or priors for functional coefficients in the regression model to improve their estimation.  The key idea is that we believe that by exploiting this regularity, we can gain more efficient and interpretable estimators, better predictions, and at times calculations that are potentially faster and more stable.  %In FDA, we wish to also exploit regularity to improve our inference.
	
Replication and regularization can be done separately using a two-step approach or together using a unified modeling approach.  For example, in functional response regression when functions are all sampled on a common grid, a two-step approach would involve fitting separate regression models for each $t$ and then smoothing the functional coefficients, or smoothing each individual function and then performing separate regression models on a grid of values for $t$.  This approach is convenient, but may be inferior to unified modeling approaches that perform replication and regularization jointly, since the borrowing of strength across functions can lead to more efficient regularization, and the regularization can more accurately inform how information is combined across functions (i.e. replication).  %This problem is reflected in the struggle to find optimal smoothing parameters in functional regression models.  
At places, this article will highlight cases in which unified modeling is done and evidence of these benefits have been shown.

Basis functions are the building blocks of FDA and typically determine the mechanism by which regularization is done. Each basis function defines a linear combination among the locations within the function, which in effect induces a correlation among those functional regions sharing loadings of high magnitude, and thus establishing a specific framework for borrowing strength. Their use also allows finite numbers of basis coefficients to yield estimates and inference in an infinite-dimensional function space.  
Some of the most commonly used basis functions in FDA are splines, Fourier series, wavelets, and principal components, and each is suited for functions with certain characteristics.  Splines are well suited to modeling smooth functions, and given their usual fitting approach work better when the dimension of the observational grid is not too high.  Fourier series transform the functions into the frequency domain, and work well for functions with stationary periodic characteristics.  Wavelets are multi-resolution basis functions that decompose a signal in a dual frequency-time domain, and their local support make them ideal for irregular functions with spikes, discontinuities, and other nonstationary features.  Both Fourier and wavelet bases possess fast algorithms for calculating basis coefficients and reconstructing the function from the basis coefficients when functions are regularly sampled on a grid of size $T$, with Fourier O($T$log$T$) and wavelets O($T$), so are well-suited for functions sampled on high-dimensional, regular grids.  Principal components (PCs) are basis functions empirically estimated from the data using eigen-decompositions.   They are well-suited for functions with simple structure  leading to rapidly decaying eigenvalues, and are able to capture distant, non-local correlations within the functions.  We will discuss some considerations and developments regarding the use of PC for FDA below.  Kernels can also be thought of as basis functions that define a weighting over which local mean estimation or regression is done, and are suitable for smooth functions when global bandwidths are used, and can be used with spatially heterogenous functions when varying bandwidths are used.  Lin et al. (2004) showed the equivalence between certain kernel and spline models.

Regularization of functions is accomplished by applying one of three operations to its basis coefficients:  [1] \textit{truncation}, [2] \textit{roughness penalties} (L2 penalization), or [3] \textit{sparsity} (L1 penalization).   For ordered bases, truncation involves the elimination of all bases after a certain point.  This is most commonly utilized in PCA, keeping only the first few PCs explaining the most variability, but can also be used with Fourier and wavelet bases ordered by increasing frequency if low-pass filters are applied, with orthogonal polynomials if polynomials above a certain order are removed, or with unpenalized regression splines for which the use of limited pre-specified knot locations can be viewed as a form of basis truncation.

Roughness penalties penalize the integrated squared second derivative of the function, which involves some type of generalized L2 penalty on the basis coefficients.  Given a single smoothing parameter, this leads to global smoothing and a ridge regression-like structure.  Fitting can be done by penalized least squares (PenLS), a Bayesian model with Gaussian priors, or a linear mixed model, which is especially convenient when using Demmler-Reinsch basis functions (Demmler and Reinsch 1975) that are orthogonalized with respect to the roughness penalty.  %This approach is commonly used with spline basis functions.  
Smoothing splines involve natural cubic spline bases with knots at each observed data point while incorporating this roughness penalty, and penalized splines involve roughness penalties but using fewer knots.  Other L2 penalties have been used in place of pure roughness penalties, with Marx \& Eilers (1996) proposing an L2 penalty on the first difference of B-spline coefficients in their \textit{P-spline} construction, and Ruppert et al. (2003) inducing a type of L2 penalization by assuming independent and identically distributed (iid) Gaussian distributions for truncated power series spline coefficients. 

 Sparsity in the basis coefficients is related to L1 penalization, which can be induced through Bayesian modeling using sparsity priors that involve stochastic search variable selection (SSVS, George \& McCulloch 1993) or nonlinear shrinkage (Bayesian LASSO, Park \& Casella 2008; Normal-Gamma, Griffin \& Brown 2010; Horseshoe, Carvahlo et al. 2010; Generalized Beta, Armagan et al. 2011), or though PenLS using L1 sparsity penalties such as the LASSO (Tibshirani 1991) or one of its many variants.  L1 penalization is a standard approach for wavelet regression, and leads to a type of adaptive regularization that removes noise while preserving dominant local features, accommodating spatial heterogeneity (Donoho and Johnstone 1995).   Sparsity is also the regularization mechanism of free-knot splines (Gervini 2006), and has been used as an alternative to truncation in PC regression when one does not want to assume that the PC dimensions explaining the most variability are necessarily the most predictive (Joliffe 1982).

Principal components play a key role in FDA and are worthy of some additional discussion.  In principal component analysis (PCA), the basis functions are empirically generated by estimating the within-function covariance surface, combining information across the replicated functions, and are thus an example of replication informing regularization.  They have been a major workhorse in FDA, especially for simple smooth functions in which case it is common for a few PCs to explain a high percentage of the total variability.  %They also are capable of capturing distant correlation between non-adjacent locations within the function.  
Care must be taken, however, in using PCA for very complex, high dimensional functional data for which the decay rate in the eigenvalues is slow, especially when the number of functions is small.  In certain high dimensional, low sample size settings PCs have been shown to be inconsistent (Hall et al. 2005; Baik \& Silverstein 2006; Jung \& Marron 2009; Lee et al. 2010; Shen et al. 2013), as in these cases there is insufficient sample size to empirically estimate the complex covariance surface characterizing the functional process.  These problems can be at least partially mitigated at times by using one of the functional principal component (fPC) approaches that incorporate regularization in the PC calculation and effectively reduce the dimensionality of the covariance surface, although it is currently not clear how to decide in a given situation whether these methods are sufficient to lead to well-estimated eigenfunctions.

Various approaches have been used to estimate regularized or sparse empirical basis functions and compute the corresponding scores using fPCA.  Two-step approaches have been proposed, whereby one first smooths the individual curves and then computes PCs, or computes the PCs and then smooths the eigenfunctions (Ramsay \& Silverman 2005, Ch 8).   Besse \& Ramsay (1986) smoothed sparse functional data to interpolate and then applied PCA.  James (2002) first smoothed the raw curves using natural cubic splines and then computed the PCs on the spline coefficients.  Johnstone \& Lu (2009) talked about sparse PCA, and described an approach that involves using sparsity in the wavelet space to obtain sparse PCA, and Mostacci et al. (2010), R\o islien \& Winje (2013), and Meyer et al. (2013) all used approaches involving wavelet denoising followed by PC calculation.  Silverman (1996) introduced a unified approach for calculating PCs while penalizing roughness.  %, and this unified approach has been demonstrated to have better properties than two-step approaches.  

Other advances in PCA have allowed the calculation of PC scores in settings with functions observed on sparse and variable grids and for multi-level models with multiple sources of stochastic variability.  In modeling repeated samples of functions within dose groups, Staniswallis \& Lee (1998) used kernels to smooth the covariance surface from which the fPCs are estimated, using quadrature to estimate the fPC scores when functions are sampled on sparse, irregular grids that vary across functions.  Yao et al. (2005a) developed a similar approach \textit{Principal Analysis by Conditional Expectation} (PACE), which borrows strength across sparse curves using a kernel-smoothed covariance matrix to estimate the eigenfunctions and computing the individual PC score as  BLUPs of a linear mixed model.  Staniswallis \& Lee (1998) performed separate fPC decompositions at two levels of a multi-level functional model, dose group and subject within dose group.   In a more general context, Di et al. (2009) presented an approach (ML-fPCA) that obtains separate fPC scores at each level of multi-level functional data based on smoothed covariance estimates constructed from method-of-moment estimates at the various levels of the multi-level models for densely sampled functions, and later generalized this to sparsely observed functions (Di, et al. 2014).  ML-fPCA was applied to estimate separate fPCs for random intercepts and slopes for longitudinally observed functional data by Greven et al. (2010), % and in various nested and crossed multi-level functional models by Shou et al. (2013).  
and Zipunnikinov et al. (2011) described how to compute the fPCs in ML-fPCA models for extremely high dimensional data, making use of the fact that $N \ll T$.  %, a strategy applied to different settings (Zipunnikinov et al. 2014).  
Bayesian (Crainiceanu \& Goldsmith 2010; Goldsmith et al. 2011b; McLean et al. 2013b) and bootstrap (Goldsmith et al. 2013) based approaches have been developed to account for the variability in PC score estimation when fitting functional linear models.  %Chen \& M\"uller (2012) performed a two-stage PC decomposition for longitudinal functional data, first estimating PCs for each of a grid of longitudinal times, and then performing a subsequent PC decomposition across the functional grid.

%Account for variability in PC estimation: Bayesian (eigenvalues, PC scores, but condition on eigenvectors; Goldsmith and Crainiceanu 2010, iterative variance/bootstrap approaches Goldsmith 2013)

%Regularization in Functional Regression:
%Purposes:  Functional predictors:
%For X(t): denies and deal with irregularly sampled predictor functions
%For B(t): reduce dimensionality of predictor to avoid collinearity, potentially improve estimation and prediction by borrowing strength, and improve interpretability.

%Purposes: Functional responses:
%1.	Account for correlation within functional errors/random effects to borrow strength, potentially gain efficiency (seemingly unrelated regression) and improve inference (more accurate coverage, joint bands, discrimination)
%2.	Avoid overfitting fixed effect functions B(t) ? improve estimation and prediction.
%Another issue is the appropriate modeling of correlation in multi-level models.  Treating correlated functions as independent can (1) lead to inefficient estimates of regression coefficients (when designs are imbalanced), and (2) inaccurate inference.  We highlight the degree to which published methods deal with this problem.
 %%% 3/19/14; 2150 words through section 2

\section{Functional Predictor Regression} \label{sec:FPR}

Most work in functional predictor regression is based on a variant of the Functional Linear Model (FLM), first introduced by Ramsay \& Dalzell (1991) and first written in its commonly encountered form by Hastie \& Mallows (1993):
\begin{eqnarray}
Y_i=B_0 + \int X_i(t) B(t) dt + E_i, \label{eq:FLM}
\end{eqnarray}
where $Y_i, i=1, \ldots, N$ is a continuous response, $X_i(t)$ a functional predictor, $B(t)$ functional coefficient, $B_0$ intercept, and $E_i \sim N(0,\sigma^2)$ residual errors.  For non-Gaussian responses, many have worked with the generalized FLM (GFLM), first introduced by Marx and Eilers (1999) for exponential family responses as
\begin{eqnarray}
g\{E(Y_i)\}=B_0+\int X_i(t) B(t) dt \label{eq:GFLM}
\end{eqnarray}  
for some link function $g(\cdot)$.

Ramsay \& Silverman (1997, Section 10.4; 2005, Section 15.3) discussed using truncated basis function expansions for $X_i(t)=\sum_{k=1}^{K_X} X^*_{ik} \phi_k(t)$ and $B(t)=\sum_{k=1}^{K_B} B^*_{k} \psi_k(t)$.  If we let $\bX_i^*=[X^*_{i1}, \ldots, X^*_{iK_X}]'$, $\bB^*=[B^*_1, \ldots, B^*_{K_B}]'$, $\bphi(t)=[\phi_1(t), \ldots, \phi_{K_X}(t)]'$, and $\bpsi(t)=[\psi_1(t), \ldots, \psi_{K_B}(t)]'$, then the functional regression terms can be written
\begin{eqnarray}
\int X_i(t) B(t) dt &=& \bX_i^* J_{\phi, \psi} \bB^*  \label{eq:basisFLM} \\
&=&  \bX_i^{**} \bB^*  \label{eq:basisFLM2},
\ee
where $\bX_i^{**}=\bX^* J_{\phi, \psi}$ and $J_{\phi,\psi}=\int \bphi(t) \bpsi(t)' dt$, and thus the FLM simplifies into a linear model of dimension $K_B$.     Notes: [1] If the same orthogonal basis functions are used for $\bphi$ and $\bpsi$, then $J_{\phi,\psi}=I$ and $\bX_i^{**}=\bX_i^*$. [2] If all $X_i(t)$ are observed on the same regular grid $\bt$ of size $T$, then one can write $J_{\phi, \psi}=\bPhi \bPsi'$, where $\bPhi=[\bphi(t_1), \ldots, \bphi(t_T)]$ and $\bPsi=[\bpsi(t_1), \ldots, \bpsi(t_T)]$.

As discussed in Section \ref{sec:basis}, regularization can be done in the basis space either through truncation, roughness penalties, or sparsity.  For functional predictor regression, this regularization can be applied to the functional coefficients $B(t)$ and/or predictors $X_i(t)$.  The regularization of $B(t)$ accomplishes several purposes: [1] reduces collinearity in the regression fit, [2] increases interpretability of coefficient estimates, [3] potentially increases estimation and prediction efficiency by making use of the functional nature of the data to borrow strength across $t$.  Best choice of basis and regularization strategy is likely to vary across data sets, depending on the characteristics of the functions and the true functional coefficient.   The use of basis functions and regularization across $X_i(t)$ has two potential purposes: [1] reduce measurement error in predictors $X_i(t)$, which for Gaussian responses increases estimation efficiency and for nonlinear models involving non-Gaussian outcomes reduces measurement-error-induced bias (Carroll et al. 1995), and [2] accommodate functional predictors with sparse and/or irregular grids that vary across sampled functions.  

%Nearly all existing methods for functional predictor regression can be written in this form, and in this paper we will link the various published methods back to this notation.

%Next show Ramsay and Silverman?s general model with basis functions for X and B ? mention regularization approaches: basis truncation and roughness penalties.  B
%The regularization of $X(t)$ has three key benefits: (1) It helps solve the collinearity problem in high-dimensional multiple regression when the functions are sampled on common fine grids.  (2) Considering X(t) measured with error, the regularization can effectively remove the measurement error from the predictor, which for linear regression has been shown to increase estimation efficiency and for nonlinear regression has been shown to remove bias and increase efficiency.  (3) When the data are observed on sparse grids that may differ across subjects, the regularization of X(t) can additionally serve as an approach to interpolate the data in a principled way to use as a functional predictor.

%The regularization of $B(t)$ has the benefit of improved estimation efficiency since strength is borrowed from nearby/correlated t in the model fitting.  Additionally, even when no basis functions are assumed for $X_i(t)$, the basis design matrix for the $B(t)$, $\Psi$, gets absorbed into the new design matrix $X_i^**=X_i \Psi$ with coefficient matrix $B^*$, which for sparse or low dimensional transforms can solve the collinearity problem. 

Most existing methodological development in functional predictor regression has followed along this general strategy laid out by Ramsay \& Silverman, with various choices of basis functions and regularization approaches.  For each method we review, we will point these out and relate their work to model (\ref{eq:basisFLM}) above.  Other model components have been introduced by others that lead to more flexible models beyond what is presented in Ramsay \& Silverman (2005), including the accommodation of non-Gaussian or correlated responses, the inclusion of other non-functional fixed or random effect terms in the model, extension to multiple functional predictors, the accommodation of functions on irregular grids, extensions to nonlinear functional predictor models, and the introduction of specific approaches for variable selection or inference.  We will highlight these developments, taking a historical context to describe the methods roughly in the order they were developed, grouped by type of basis functions used.

%Describe historical development of multivariate approaches involving PCR/PLS/RR.  Mention don?t explicitly account for functional nature of data.  Mention historical development:

\textbf{PCA for $X_i(t)$ and $B(t)$:}	The original work on FLM was done using multivariate analysis techniques for functions sampled on a common grid, with principal component regression (PCR, Kendall 1957), partial least squares (PLS, Wold 1966), or ridge regression used to reduce collinearity in the resulting high dimensional multiple regression.   In PCR, orthogonal empirically determined principal component bases estimated from the decomposition of $X'X$ are used for both $\bPhi$ and $\bPsi$.  According to Joliffe (1982), the original work on PCR (e.g. Kendall 1957) suggested a variable selection approach to regularization, supposing that the PCs explaining the largest amount of variability in $X_i(t)$ may not necessarily be the most important in predicting $Y_i$.  However, since in many cases the first few PCs appear most important, over time many researchers have regularized by truncating at the first few PCs explaining a fixed proportion of total variability.  Both strategies have been used in practice.  Other methods extending PCR have been developed for function predictor regression.  %Unfortunately, none of these classical multivariate approaches take the ordering inherent in the functional data into account.  In discussion of an article on PCR, PLS, and RR, 

Cardot et al.  (1999) fit a FLM with PCR through truncation and discussed some theoretical results.   M\"uller \& Stadtm\"uller (2005) introduced functional PCR (fPCR) for GFLM, using smoothed fPC bases for both $X_i(t)$ and $B(t)$ with regularization done by PC truncation.  The basis functions were estimated by decomposing a kernel-smoothed estimate of the covariance function after subtracting off a kernel-smoothed estimate of the overall mean, while removing white noise errors.   When the functions are irregularly and sparsely sampled, the PC scores $Y^*_i$ are computed using Principal Analysis by Conditional Expectation (PACE, Yao et al. 2005a), which borrows strength across sparse curves to estimate the covariance matrix and estimates individual PC scores as BLUPs of a linear mixed model.  %Kong et al.  (2014) developed asymptotic hypothesis tests for FLM based on fPCR for Gaussian responses and regular or irregularly sampled functions.  

In a multi-level functional data setting with multiple curves per subject, Crainiceanu et al.  (2009)  used ML-fPCA (Di et al. 2009) %, which uses method-of-moments to estimate random effects and residuals in a multi-level model, and then performs a separate fPC decomposition at each level
 to estimate the multi-level PC scores, using a truncated set of subject-level scores as the predictors $X_i^{**}$ in model (\ref{eq:basisFLM2}),  %akin to PACE (Yao, M\"uller, and Wang 2005).  A truncated set of subject-level PCs are then used as predictors in a linear model, 
effectively using subject-level PCs as bases for both $\bphi$ and $\bPsi$ and regularizing by truncation.   They used RLRT (Crainiceanu \& Ruppert 2004) to select the truncation parameter, and also discussed a fully Bayesian MCMC-based version to obtain estimates and inference.  Crainiceanu \& Goldsmith (2010) developed a Bayesian modeling approach for PCR using WinBugs, modeling the PC scores and eigenvalues stochastically (but still conditioning on estimated eigenvectors as fixed), which they showed leads to more accurate inference.

Holan et al. (2010) developed Bayesian methods to classify subjects based on a functional predictor that is a nonstationary time series.  They computed the spectrogram, a time-varying Fourier spectrum, to transform the time series to the time-frequency dual space, and then constructed a logistic GFLM with the spectrogram image as a predictor, using PC for the basis functions and SSVS for regularization.  
Randolph et al. (2012) introduced a method for PCR for which regularization is done by a structured L2 penalty that incorporates presumed structure directly into the estimation process through a linear penalty operator, leading to a weighted ridge regression in the PC space.

\textbf{Splines for $X_i(t)$ and/or $B(t)$:} Hastie \& Mallows (1993) pointed out that standard multivariate approaches like PCR do not take the ordering inherent to the functional data into account and mentioned the growing importance of functional data modeling.  They introduced the FLM (\ref{eq:FLM}), and used penalized splines for $B(t)$ while assuming all functions were sampled on a common grid $\bt$.  Others have also incorporated spline bases for regularization in the FLM or GFLM settings.%Note that although no basis is explicitly assumed for $X_i(t)$, the model simplifies into a penalized least squares regression where the predictors are effectively the spline coefficients for $X_i(t)$, so dimension reduction and reduction of collinearity is obtained.  $X_i^{**}=X_i \Psi$ and $B^{*}$ are spline coefficients.
	
Marx \& Eilers(1999) introduced \textit{penalized signal regression} (PSR) for the GFLM %, designed for functions sampled on a regular, common grid, and using 
that uses B-splines for $B(t)$ that are penalized by the \textit{P-spline} first difference L2 penalty introduced in Eilers and Marx (1996).  %approach involving the first difference penalty of Eilers and Marx (1996).  % regularizing via a first difference penalty introduced by Eilers and Marx (1996) that approximates the formal 2nd derivative smoothness penalty.  
Designed for functional predictors on a common grid $\bt$, their approach does not transform $X$ ($\bPhi=I_T$) but uses the B-spline design matrix for $\bPsi$, resulting in $X^{**}=X \bPsi'$ in (\ref{eq:basisFLM2}).   They subsequently extended this work in various ways, adding multiple linear and additive scalar fixed effects and linear random effects to the model (GLASS, Eilers \& Marx 2002), accommodating multidimensional signals such as images (MPSR, Marx \& Eilers 2005), and allowing prespecified weights for adaptive smoothing (SPSR, Li \& Marx 2008).  %All of their work assumes each function is observed on the same common equally-spaced grid $\bt$.
	
James (2002) fit a GFLM using natural cubic spline bases to represent the predictor functions $X_i(t)$ with regularization done by the truncation inherent in the knot selection.  To account for measurement error and accommodate varying sampling grids across functions, they specified a measurement error model $X_i=X_i^* \bPhi_i + \epsilon_i$, where $X_i$ and $\bPhi_i$ are vectors with the functional predictors and natural cubic spline basis functions evaluated on the observational grid $\bt_i$, and $X_i^*$ a vector containing the spline coefficients for a common set of knots.

	Zhang et al. (2007) presented a spline-based FLM with periodic spline bases used for both $\bPhi$ and $\bPsi$, and regularization by roughness penalties.  Like James (2002) and M\"uller \& Stadtm\"uller (2005), their approach accommodated irregularly sampled functions and adjusted for measurement error in $X_i(t)$.  %After estimating the spline coefficient of the denoised functional predictor, they fit using linear mixed models, and presented a score test for testing the functional coefficient alternative vs. constant null.  
Crambes et al. (2009) presented a smoothing-spline based FLM, and proved some theoretical results, and Maronna \& Yohai (2013) presented a robust version of this model that is insensitive to outliers.  Yuan and Cai (2010) introduced a general reproducing kernel Hilbert space approach to functional linear regression regularized by roughness penalties, implemented using smoothing spline-like kernels.
	
\textbf{Wavelets for $X_i(t)$ and $B(t)$:}	
	Although not explicitly describing their model as functional regression, Brown et al. (2001) introduced a Bayesian, wavelet-based approach for fitting FLMs to data on a common equally-spaced grid $\bt$.  They used orthogonal wavelets bases to represent both $X_i(t)$ and $B(t)$, and used SSVS to select a subset of important wavelet coefficients.   Their approach effectively used the orthonormal wavelet transform matrix for both $\bPhi$ and $\bPsi$, but coefficients were actually calculated using the O($T$) pyramid-based discrete wavelet transform (DWT) algorithm, allowing the approach to scale up to functional data on very large grids.  As mentioned above, the use of wavelets and sparsity leads to adaptive regularization, making it well-suited for modeling spatially heterogeneous functions with local features like spikes.  % were both assumed to be the orthonormal DWT matrix, but one of the key benefits of wavelets is that fast algorithms are available to compute the coefficients in O($T$) time without matrix multiplication or least squares, making them useful for  enormous data sets.  
%Additionally, wavelets have the advantage of being able to adaptively regularize the $B(t)$ function, avoiding the attenuation of local features in the curve that characterizes non-adaptive smoothing approaches, and making them especially suitable for spiky predictors and coefficient functions.  
Subsequent work involving wavelet representations for $X_i(t)$ and $B(t)$ include Wang et al. (2007), Malloy et al. (2010), and Zhao et al. (2012).  Wang et al. (2007) predicted a dichotomous response using a Bayesian probit regression model, first denoising the predictor function $X_i(t)$ using a Bayesian LASSO to induce L1 penalization on the wavelet coefficients, and then performing SSVS on these coefficients to adaptively regularize $B(t)$.  Malloy et al. (2010) fit lag time models using wavelets with regularization done by SSVS, and included scalar fixed and random effect predictors to model correlated responses from a nested design.  Zhao et al. (2012) modeled using wavelets and regularized using the LASSO as a sparsity prior. %An alternative approach for handling spiky predictors is presented in Woodard, Crainiceanu, and Ruppert (2013).  Using kernel mixture functional representations for the spiky functions, they effectively extract features from the function selecting from an overcomplete basis, and then use as predictors in a second-stage regression of a hierarchical Bayesian model.  The regression model involves smooth additive terms for each feature, parameterized by B-splines regularized by truncation, and SSVS is used to select among the additive feature predictors.  
	
\textbf{General Bases for $X_i(t)$ and/or $B(t)$:} 
	Ratliffe et al. (2002a) presented methods for the FLM using common basis transform $\bPhi$ for $X_i(t)$ and $B(t)$, with $X_i^{**}=X_i^{*} \bPhi \bPhi'$ and regularization done by truncation.  They used Fourier basis coefficients, but presented their method for any general basis, extending to a logistic FLM in Ratliffe et al. (2002b).  Ogden et al. (2002) transformed 2D image data into 1D variance functions, to which they fit a FLM using Fourier Bases for both $X_i(t)$ and $B(t)$, regularizing by roughness penalty and fitting by PenLS.
	
	Zhu \& Cox (2009) developed a group-lasso approach for performing variable selection across multiple potential functional predictors in a GFLM.  Assuming regularly sampled functional predictors, they assumed a common truncated orthonormal basis transform for both $X_i(t)$ and $B(t)$.
	Zhu \& Vannucci (2010) and Lee \& Park (2011) utilized sparsity in the basis coefficients for regularization.  Zhu \& Vannucci (2010) modeled dichotomous outcomes using a Bayesian probit model, using general orthogonal basis functions for both $X_i(t)$ and $B(t)$, truncating at a large number and then performing SSVS to select among them. Lee \& Park (2011) introduced a FLM for Gaussian responses that utilizes a common general basis transform for both $X_i(t)$ and $B(t)$ and regularizes via one of a number of sparsity penalties on the coefficients, including LASSO, adaptive LASSO, or SCAD.
Fan \& James (2014) also used the group-lasso to perform variable selection across multiple functional predictors, using a common orthogonal basis transform for $X_i(t)$ and $B(t)$, in their case for Gaussian responses.
	
James et al. (2009) introduced a method (FLiRTI) that encourages sparsity in the $B(t)$ for interpretability, accomplished via L1 penalization to shrink the coefficients towards sparsity in a prespecified derivative.  They used piecewise constant basis functions for $B(t)$, but mentioned their approach can be used for any high-dimensional basis expansion including splines, Fourier, and wavelets. 	

Goldsmith et al. (2014) introduced a Bayesian method to predict a Gaussian response from an image predictor, using a combination of Ising prior and Markov Random Field prior to encourage significant regions of the the coefficient surface to cluster and borrow strength from each other.

\textbf{Combinations of PCs and Splines for $X_i(t)$ and $B(t)$:}	A number of papers have used PCs and splines together to perform regularization.  
One approach has been to represent the functional predictor $X_i(t)$ using splines, and then perform a PC decomposition of the spline coefficients.  After fitting unpenalized cubic splines to the functional predictors, James (2002) performed a principal component decomposition on the matrix of spline coefficients $X^*=[X_1^*, X_2^*,\ldots, X_N^*]$, and used this basis to further transform the $X(t)$.   %Regularization of $B(t)$ was done by truncation of the PC bases. 
 In the form of (2), they effectively used natural spline bases for $\bPhi$ and eigenvectors of the spline bases for $\bPsi$, and fit their model using the EM algorithm and weighted least squares, with the spline coefficient eigenvalues as the weights.   

Reiss \& Ogden (2007)  used a similar strategy in a method that combines together the ideas of PCR and PSR (Marx \& Eilers 1999).  For Gaussian outcomes, their model effectively transforms $X_i(t)$ using a $T \times K_x$ B-spline design matrix $\bPhi$, and then uses a PC decomposition on $\mathbf{X} \bPhi$ to estimate a truncated set of eigenfunctions $\bPsi$, with $X^{**}=X \bPhi \bPsi$ in (\ref{eq:basisFLM2}).  Regularization is done via roughness penalties (or P-spline penalties) applied to either the fPC (fPCR$_C$) or the regression coefficients (fPCR$_R$).  They presented a similar strategy but replacing PC bases with PLS bases (fPLS$_C$, fPLS$_R$), empirical bases that maximize correlation between $X_i(t)$ and $Y_i$ instead of simply explaining variability in $X_i(t)$.  Reiss \& Ogden (2010) extended the fPCR$_R$ approach to handle exponential family responses and 2D image predictors.  Radial basis B-splines were used to ensure radial symmetry in the images.  They also presented inferential methods, including bootstrap-based simultaneous confidence bands and using RLRT (Crainiceanu \& Ruppert 2004) to test the functional coefficient alternative vs. a constant null.   
	
An alternative approach is to transform the functional predictors $X_i(t)$ using PCA while parameterizing the coefficient function $B(t)$ using splines.  Cardot et al. (2003) introduced a 2-step method that performs PCR and then uses B-splines with a roughness penalty to smooth the resulting $B(t)$.	  Crainiceanu and Goldsmith (2010) presented a Bayesian GFLM with PCs for $X_i(t)$ and B-splines for $B(t)$, regularizing using the random walk prior of Lang \& Brezger (2004) on the spline coefficients, and modeling the PC scores and eigenvalues stochastically.   Goldsmith et al. (2011b) developed a variational Bayes (VB) approximation for this model, and found this was sufficient for estimation but not necessarily for inference.  Goldsmith et al. (2011a) described a frequentist variant of this method called penalized functional regression (PFR) using PCs for $\bPhi$, truncated power series for $\bPsi$, and regularizing by L2 penalties on the truncated power series spline coefficients as in Ruppert et al. (2003).   Their approach can handle sparsely observed, irregular, or multi-level functional data using PACE or ML-fPCA to obtain PC scores to use in the GFLM as predictors, and also includes non-functional linear predictors.  Instead of truncation they suggest keeping a large number of PCs, making the primary purpose of their PC decomposition the handling of irregularly sampled functions rather than dimension reduction.   Note that when used with regularly sampled data and keeping all PCs, their approach is like PSR (Marx\& Eilers, 1999) but using truncated power series and iid Gaussian L2 penalties in place of the B-splines with first difference penalties.  Swihart et al. (2013) presented functional hypothesis testing procedures based on RLRT for comparing functional vs. constant effects and selecting among multiple functional predictors in this setting.  

Goldsmith et al. (2012) extended this approach to handle data with repeated measurements in both the response and functional predictor by adding scalar random effects to the GFLM.  Their specific model includes iid random effects (e.g. to model nested data), but clearly could be applied using other more complex random effects structures, as well.  Their approach does not account for the correlation among replicate functions in the calculation of the PC scores, but Gertheiss et al. (2013) presented an alternative approach that does.  Using an extension of ML-fPCA (Di et al. 2009) for longitudinal functional data (LFPCA, Greven et al. 2010), they computed separate eigenvectors for random intercepts and slopes, estimated a shared PC score between them using a regression approach, and then carried this PC score forward into the GFLM.   

\textbf{Nonlinear functional predictor models:} The previously described methods all assumed linear functional effects.  A number of methods for nonlinear functional predictor regression have been proposed.

Li \& Marx (2008) extended the PSR method of Marx and Eilers (1999) to include general additive polynomial terms like $\int \{X_i(t)\}^2 B_2(t) dt$.  Yao \& M\"uller (2010) allowed full quadratic functional predictors, 
\begin{eqnarray}
Y_i= B_0 + \int X_i(t) B_1(t) dt + \int \int X_i(t) X_i(s) B_2(t,s) dt ds + E_i.
\end{eqnarray}
Note that this yields greater flexibility than Li \& Marx (2008) which only allows the diagonal cross product in the second term.
They used PC decompositions to estimate empirical basis functions for $X_i(t)$, and used these as basis functions for $X_i(t)$, $B_1(t)$, and $B_2(t,s)$, with regularization by truncation.  
Their approach generalizes to full polynomials of any order.  Yang et al. (2013) extended the spectrogram-based method of Holan et al. (2010) to 
this penalized quadratic regression setting, using a Bayesian modeling approach with SSVS across the PC dimensions for regularization.   	

In the first nonparametric extension of the FLM, James \& Silverman (2005) introduced a method Functional Adaptive Model Estimation (FAME) that extends projection pursuit regression to functional predictors and exponential family outcomes.  The model is given by
\begin{eqnarray}
	g\{E(Y_i)\}=B_0+\sum_{k=1}^K f_k\left\{\int X_i(t) B_k(t)\right\} dt,
\end{eqnarray} 
where $f_k$ is a smooth function of unspecified form.  % and for identifiability, they assumed $\int B_k(t) dt=1$ and corr[$f_k\{\int X_i(t) B_k(t) dt\}, f_{k'}\{\int X_i(t) B_{k'}(t)dt\}$]=0.  
They used natural cubic splines to represent $X_i(t)$, $B_k(t)$, and $f_k(\cdot)$, and regularized using roughness penalties.  Their model was extended to included multiple functional predictors.
	
Various researchers have incorporated nonlinearities in functional predictor regression using a special case of projection pursuit, single index models, involving $f(\int X_i(t) B(t) dt)$ for some smooth $f(\cdot)$.  Li et al. (2010) used single-index models to model interactions of scalar and functional predictors, using complete orthogonal basis representations for the $X_i(t)$.  Marx et al. (2011) used single-index models to extend the PSR of Marx \& Eilers (1999) in the setting of 2d functional predictors, using tensor B-splines for the predictor surface and 1d B-splines for the index function, regularizing using P-spline penalties.   Fan \& James (2014) allowed separate additive single index terms for $p$ functional predictors, using a common orthogonal basis for $X_{ij}(t)$ and $B_j(t)$ and another basis for $g_j(\cdot)$, $j=1,\ldots,p$, using truncation for penalization and group lasso to perform variable selection across the functional predictors. 
 	
M\"uller \& Yao (2008) introduced Functional Additive Models (FAM) which extend the fPC regression approach of Yao et al. (2005)  to nonlinear models that involve additive nonparametric functions of the PC scores, using the model
\begin{eqnarray}
	Y_i = B_0 + \sum_{k=1}^{K_x} f_k(Y^*_{ik}) + E_i,
\end{eqnarray}
where $Y^*_{ik}$ are the PC scores for a truncated set of $K_x$ PCs, and $f_k$ a general function smoothed using kernel approaches.
Zhu et al. (2014) introduced a variant of this approach that did not perform truncation over PCs, but instead used COSSO regularization, a form of L1 penalization, based on an assumption that the function lives in a particular reproducing kernel Hilbert space.  This allows dimensions explaining less total variability in $X_i(t)$ but more important for predicting $Y_i$ to be included in the regression.
% Wu et al. (2010) fit varying coefficient FLMs (VC-FLM) which allows the functional predictor to vary as a function of another covariate $V$, $B(V,t)$
	
McLean et al. (2012) presented a functional generalized additive model (FGAM) that extends generalized additive models (GAM, Hastie \& Tibshirani 1986) to the generalized functional predictor regression framework for noise-free functions observed on a common grid $\bt$:
	\begin{eqnarray}
g\{E(Y_i)\} = B_0 + \int f\{X_i(t), t\} dt + E_i(t),
	\end{eqnarray}
where $g$ is a general link function for an exponential family and $f(x,t)$ is a smooth functional additive regression surface.  They parameterized this 2d surface using tensor B-splines, regularized by P-spline-type L2 penalization, and fit the model by penalized iteratively weighted least squares with generalized cross validation (GCV) to estimate the smoothing parameters.  McLean et al. (2013) introduced a Bayesian FGAM for sparsely observed functions on irregular grids.  Effectively extending Crainiceanu and Goldsmith (2010), they handled the measurement error and unequally spaced $X_i(t)$ using PC decompositions, updating the PC scores within the MCMC conditional on the eigenfunctions, eigenvalues, and mean curves.  They used tensor product B-splines with random walk penalties on the coefficients as in Lang \& Brezger (2004).   %M\"uller et al. (2013) also introduced an additive model for exponential family functional predictor regression, using orthonormal basis functions for both the functional domain $t$ and additive predictor $x$, with regularization by truncation.
	
\textbf{Summary of Functional Predictor Regression Models:}  There has been a great deal of work in functional predictor regression, especially in the past 10 years.  While some methods include nonlinear effects of some sort, most methods work with a version of the Gaussian or generalized functional linear model, (\ref{eq:FLM}) or (\ref{eq:GFLM}), respectively.  They key differences among the methods are their choice of basis to represent the predictors $X_i(t)$ and/or the functional coefficient $B(t)$, with the most common choices being principal components, splines, wavelets, or some combination thereof, and their approach for regularization.  No single basis function approach is likely to be superior in all settings, as each makes sense for particular types of data, as discussed in Section \ref{sec:basis}.  New methods for choosing among various potential basis functions and assessing which regularization approaches work best would be a welcome addition to the literature, and help users assess which approach to use given a particular data set.

\textbf{Available Software for Functional Predictor Regression:}  A number of the methods discussed in this section have publicly available software.  The methods from Ramsay and Silverman (2005)'s textbook are contained in the R package $fda$ and are also available in Matlab, as described in an accompanying computational textbook (Ramsay, et al. 2009).  The R package $fpca$ implements the sparse PCA approaches of James, et al. (2000), and R and Matlab code for James and Hastie (2001) and FLiRTI (James, et al. 2009) are available at \url{http://www-bcf.usc.edu/~gareth/research/Research.html}.  The PACE software (Yao, et al. 2005a) and its applications can be downloaded at \url{http://www.stat.ucdavis.edu/PACE/}, and the software from M\"uller and Stadtmuller (2005) are available at \url{http://anson.ucdavis.edu/~mueller/data/spqr.html}.  The software for a number of methods can be found in the $refund$ R package which started with $fpcr$ (Reiss and Ogden 2007, 2010) and later was appended with the packages $pfr$ (Goldsmith, et al. 2011), $lpfr$ (Goldsmith et al. 2012), $peer$ (Randolph, et al. 2012), $lpeer$ (Kundu, et al. 2012), $wnet$ (Zhao, et al. 2012), and $fgam$ (McLean et al. 2013).  The software for the Bayesian methods of Goldsmith, et al. (2010) can be found at \url{http://www.biostat.jhsph.edu/~ccrainic/webpage/software/Bayes_FDA.zip}.

\section{Functional Response Regression} \label{sec:FRR}

Functional response regression involves the regression of functional responses on a set of scalar predictors.  Given a sample of functional responses $Y_i(t_j), i=1, \ldots, N; j=1, \ldots, T_i$ and scalar predictors $X_{ia}, a=1,\ldots,p$, a general linear functional response regression model is given by
\begin{eqnarray}
Y_i(t_j)=\sum_{a=1}^p X_{ia} B_a(t_j) + E_i(t_j), \label{eq:FRR1}
\end{eqnarray}
where \textit{functional coefficient} $B_a(t)$ represents the partial effect of predictor $X_a$ on the response at position $t$.  The goal of functional response regression is often estimation of $B_a(t)$ and then either testing whether $B_a(t)=0$ or assessing for which $t$ is it true that $B_a(t)\ne 0$.  The $E_i(t)$ are the curve-to-curve residual error deviations, frequently assumed to be iid mean zero Gaussians with covariance $S(t_1, t_2)$, whose structure describes the within-function covariance.  Sometimes the curve-to-curve deviations are split into the sum of a \textit{curve-level random effect function} $U_i(t_j)$ and white noise residual errors $\epsilon_{ij} \sim \mathcal{N}(0,\sigma^2)$, with the primary benefit being the ability to obtain denoised estimates of the individual curves as $\hat{Y}_i(t)=\sum_a X_{ia} \hat{B}_a(t) + \hat{U}_i(t)$.   If these random effect functions are represented by a truncated basis representation $U_i(t)=\sum_{k=1}^{K_u} U_{ik}^* \phi_k(t)$, and it is assumed that $\bU^*_{i} \sim \mathcal{N}(0,S^*)$ where $\bU^*_{i}=(U^*_{i1}, \ldots, U^*_{iK_u})'$ and $S^*$ is a $K_u \times K_u$ covariance matrix for the basis coefficients, then this implies that $S(t_1, t_2) = \boldsymbol{\phi}(t_1) S^* \boldsymbol{\phi}(t_2)'+\sigma^2 I(t_1=t_2)$ , where $\boldsymbol{\phi}(t)=(\phi_1(t), \ldots, \phi_{K_u}(t))$.  Thus, the form of the within-function covariance is determined by choice of basis and basis space covariance matrix $S^*$.  

%The primary goals in functional response regression include estimating the functional coefficients, performing functional or point-wise inference to assess whether $B_a(t)=0$ or to determine the set of $t$ such that $B_a(t) \ne 0$, or can be used to perform functional discriminant analysis, as described below.

The concepts of replication and regularization enter into functional response regression modeling in various ways.  Replication is involved in the regression, and also in accounting for potential between-function correlation, % through \textit{random effect functions} in \textit{functional mixed effect models}, 
which will be discussed later in this section.   Regularization of the functional coefficients $B_a(t)$ increases interpretability,  potentially increases estimation and precision accuracy, and allows interpolation to values of $t$ between sampled grid points, and regularization of the curve-level random effect functions $U_i(t)$ can have similar benefits.  As with functional predictor models, this regularization is typically done using either truncation, roughness penalties, or sparsity in some chosen basis space.   Further, the inclusion of some within-function correlation structure in $S(t_1, t_2)$ induces regularization, as it defines a manner by which strength is borrowed within the function in estimation and inference.  %which in turn affects estimation and inference of the functional coefficients $B_a(t)$.  
This affects estimation of $B_a(t)$ when taken into account using weighted least squares (WLS) for which the within-function covariance assumptions determine the weights (Hart \& Wehrly 1986;  Lin et al. 2004; Wu \& Liang 2004; Morris \& Carroll 2006; Krafty et al. 2008; Reiss et al. 2010; Staicu 2010), as is typical for proper regression analyses of correlated data.  This borrowing of strength is related to the idea of seemingly unrelated regression (Zellner 1962).  Further, inference on $B_a(t)$ is affected by the within-function covariance assumptions, especially joint inference such as confidence bands (Lee \& Morris, 2014).

Much of the existing work on functional response regression can be related back to model (\ref{eq:FRR1}).  Methods differ in the scope of their models, their mode of regularization of $B_a(t)$, and the assumptions made on the within-function covariance $S(t_1, t_2)$.  % the form of regularization done on the functional coefficients and the within-function covariance assumptions made in $S(t_1, t_2)$, and models vary in the %and there are clusters of work that focus on a special case of this model where the functional coefficients are limited to a single overall mean (growth curves) or several group means (functional ANOVA).  
We will highlight these aspects of each method, focusing on the modeling, but sometimes with brief discussions of inferential capabilities of the models if they are emphasized.  Some methods work for sparse, variable grids between functions, while others require regular, fine, and/or common grids across functions.  Some approaches are able to model correlations between functions, and for these methods we will highlight their capabilities and manner of capturing this correlation.  Most methods in the literature are for 1d functions on Euclidean domains and involve linear modeling (in $X$), but some methods handle 2d functions like images, functions on non-Euclidean manifolds, or allow nonlinearities (in $X$), and these will also be highlighted.  Most methods are designed for functions sampled on small to moderate sized grids, although some have been designed to scale up to enormously sized functional data, e.g. the \textit{mass spectrometry data set} in Section \ref{sec:data}.

\textbf{Growth Curves:}  Early work focused on estimation of the mean function $ \mu(t)$ ($X_{ia}=\mathbf{1}$) from an iid sample of curves, which has been called \textit{growth curve analysis} in the classic setting where $t$ represents time.  Rao (1958) talked about the use of PCs to represent $\phi(t)$ in the curve-to-curve deviations, in which case the form of $S$ is determined by the estimated PCs.  Laird \& Ware (1982) discussed linear mixed models (LMM), which can directly be used to fit growth curves when linear parametric forms are used for the mean function and curve-to-curve deviations, frequently through  orthogonal polynomials. Hart \& Wehrly (1986) nonparametrically estimated the mean of a sample of curves using kernel smoothing with parametric assumptions for $S$, e.g. an AR(1).  They found that the optimal bandwidth could be smaller or larger depending on the autocorrelation, demonstrating that the within-function covariance assumptions affect the regularization of the functional mean estimation.  Rice \& Silverman (1991) estimated the mean of a sample of curves nonparametrically using splines with roughness penalties, with $S$ parameterized using a fPC decomposition involving a roughness penalty in PC calculation, with regularization by truncation.
 
 Barry (1995) introduced a Bayesian model with Gaussian processes for both the overall mean curves and curve-to-curve deviations, which is like a smoothing spline with separate roughness penalties for the mean curves and deviations.  Shi et al. (1996) represented the mean curve with B-splines regularized by truncation inherent in the knot selection, and $S$ parameterized by a PC decomposition of the B-spline coefficients plus white noise.  Zhang et al. (2000) proposed a semi-parametric model that adds parametric fixed effects to a model with a nonparametric mean function, using periodic spline bases for the mean curve, regularizing by roughness penalty.  They represented $S$ parametrically using an Ornstein-Uhlenbeck process, and performed global testing for the difference of two mean curves.  
 
 James et al. (2000) represented the mean curve using  B-splines with regularization by truncation inherent in the knot location, represented the curve-to-curve deviations with B-splines, modeled the basis space covariance matrix $S^*$ as unstructured or through PC decompositions, and included white noise errors.   Martinez et al. (2010) implemented a Bayesian version of this approach using roughness penalties for the mean curve and curve-to-curve deviations, and using reversible jump MCMC (RJMCMC) to perform Bayesian model averaging over the number of PCs in the model.  Rice \& Wu (2001) also used regression splines with fixed knots to represent the overall mean function and curve-to-curve deviations, both regularized by the truncation inherent in knot placement, and the covariance $S^*$ left unstructured.
  
 Wu \& Zhang (2002) used local polynomials to represent the mean function and curve-level random effect functions, with independent but heteroscedastic residual errors.  Yao et al.  (2005a) kernel-smoothed the overall mean function, and then represented $S$ using fPC obtained by performing a PC decomposition of the kernel-smoothed covariance matrix, after subtracting off white noise errors.  For functions sampled on a common grid, Gervini (2006) represented curve-level random effect functions using a PC decomposition, with the fPCs and overall mean function both adaptively regularized using free-knot splines.
 
 Bigelow \& Dunson (2007) presented a Bayesian approach fitting the overall mean and curve-level random effect functions using truncated linear splines, regularizing using sparsity priors at knot locations, with the between-spline-coefficient covariance $S^*$ assumed diagonal but heteroscedastic, and using RJMCMC for model fitting.  Thompson \& Rosen (2008) also presented a Bayesian approach representing the overall mean and curve-level random effect functions using B-splines, regularized using sparsity priors applied to large set of candidate knots, and with $S^*$ left unstructured with an inverse Wishart prior.  Fox \& Dunson (2012) presented a Bayesian approach that represents the mean function and curve-level random effects using multi-resolution Gaussian processes, adaptively regularized through SSVS across potential multi-resolution partitions.  $S$ was constrained to be the sum of multi-resolution piecewise, weighted exponential kernels plus white noise.  Storlie et al. (2013) presented a Bayesian approach representing the mean and curve-specific functions as Gaussian Processes using a covariance kernel based on Bernoulli polynomials that has connections to smoothing splines, with this kernel determining the structure of $S$ up to four scalar variance components.
 
 Ogden \& Green (2010) projected the observed curves into the wavelet space to estimate mean functions for curves sampled on a fine, common grid.  The mean function was regularized by inducing sparsity in the wavelet space through multiplicity-adjusted hypothesis tests for each wavelet coefficient, and, like in Morris \& Carroll (2006), the wavelet-space covariance $S^*$ was diagonal but heteroscedastic across wavelet coefficients, which leads to nonstationary features in $S$, allowing heteroscedasticity and varying degrees of smoothness across $t$.  %given multi-resolution wavelet bases with local support leads to a broad class of induced nonstationary covariance structures for $S$, accommodating heteroscedasticity and varying degrees of smoothness across $t$.   
 Similarly, Bunea et al. (2011) also modeled functional data on a fine, common grid and used a flexible sparse basis like wavelets to represent the mean function, regularizing via sparsity achieved by hard thresholding rules.  Staicu et al. (2012) introduced methods for estimating the mean curve for functions with skewed distributions, using a parametric distribution with shape and scale parameters varying over $t$, using $t$-copulas to account for within-function correlation.
 
%Staicu et al. (2012) introduced methods for modeling growth curves with skewed distributions.  For the residual errors, they assumed some parametric distribution with shape and scale parameters allowed to vary over $t$.  They used a multi-step method for estimation, first undersmoothing the raw curves obtain data on a common grid, estimating the mean, standard deviation, and shape point wise for each $t$ using penalized maximum likelihood, and then smoothing while accounting for the within-function correlation through a $t$-copula.  

\textbf{Joint Bands and Multiple Testing:}  Various researchers have proposed joint confidence/credible bands with global functional coverage for the mean functions in growth curve models.  Li \& Hsing (2010) and Degras (2011) dealt with densely sampled functions and used local linear kernel approaches, and Bunea et al. (2011) used orthonormal basis functions.  Ma et al. (2012) dealt with sparsely sampled functions and used piecewise constant splines.  Storlie et al. (2013) presented approaches using B-splines and Gaussian processes.  Using an approach described in Ruppert et al. (2003), joint credible bands can be constructed from MCMC samples of any Bayesian functional model or from bootstrap samples based on a Gaussian approximation.  Crainiceanu et al. (2012) used this approach to produce bootstrap-based joint bands for the difference of two mean functions, considering various parametric and nonparametric bootstrap strategies and regularizing point-wise estimates of the mean functions through penalized splines.  This approach has also been used for joint inference on functional regression coefficients in the general Bayesian functional mixed model framework (Meyer et al. 2013; Lee and Morris 2014; Zhang, et al. 2014; Lee, et al. 2014).  Joint bands can be inverted to yield functional inference (e.g. see simultaneous band scores, \textit{SimBaS}, in Meyer et al. 2013) that controls global type I error rate.  Additionally, for Bayesian methods posterior probabilities can be computed that have connections to local false discovery rate (e.g. see Morris et al. 2008), and can be used to flag regions of functional regression coefficients while controlling FDR.

\textbf{Functional ANOVA:} Some papers have generalized growth curves to multiple populations ($X_{ia}=\delta_a(i)$), which can be viewed as functional regression with discrete predictors, or functional ANOVA.  Barry (1996) extended previous work to test for differences in mean curves, using random intercepts to induce compound symmetry structure in $S$.  For growth curves sampled within each of a number of doses, Staniswallis \& Lee (1998) estimated mean functions across a set of doses using a complete, orthonormal basis expansion in $t$, represented as the inner product of the basis functions and separate nonparametric smooth functions of dose for each basis function.  They represented $S$ using fPCs, with an eigendecomposition done on kernel-smoothed estimates of the covariance matrices, plus white noise errors, and used Bonferroni-corrected ANOVA tests to perform inference.   Spitzner et al. (2003) constructed FANOVA tests using PCs or Fourier bases to represent 2d functional responses, and Abramovich \& Angelini (2006) constructed FANOVA tests based on wavelet representations of the functions.

%For EEG data, Wang et al. (2009) presented a functional response regression model with two-way ANOVA fixed structure, a single level of random effect functions at the curve level, and iid errors.  Given functions sampled on a common grid, they consider a two-step method similar to Fan \& Zhang (2000) first fitting linear mixed models to each $t$ and then smoothing the functional estimates, and also considered a basis representation approach, where different basis functions (typically B-splines) are used to represent $Y(t)$, $B(t)$, and $U(t)$, and absorbed into the design matrix.  Their random effects covariance in the basis space is given by general matrix $S^*$, which along with the choice of basis functions determines the within-function correlation, and fixed effects are regularized by choice of basis, i.e. truncation.  They fit using PROC MIXED and perform functional ANOVA.

\textbf{General Functional Response Regression:}  Many of the methods for functional growth curves and functional ANOVA described above could in principle be generalized to the functional response regression setting, regressing the function on a general design matrix $\mathbf{X}$ as in model (\ref{eq:FRR1}), with some additional software development required.  Some researchers have developed methods with the general model (\ref{eq:FRR1}) specifically in mind.  Faraway (1997) described a two-step approach in which they smooth or interpolate the observed curves, and then perform point-wise regressions for each $t$, ignoring the within-function correlation, and then performing inference using a bootstrap or multivariate techniques based on an eigendecomposition of the empirically estimated curve-to-curve deviations.   Ramsay \& Silverman (1997)  first proposed separate point-wise regressions for each $t$ with no regularization, %and mentioned that even without regularization the functional coefficients are typically smoother than original $Y_i(t)$.  %They suggested that regularization is simply cosmetic in this case, although other evidence suggests this is not true.  
and second discussed regularizing the functional regression coefficients through roughness penalties and basis function modeling.  Their focus was estimation not inference, and they assumed white noise errors and no curve-level random effects, so effectively used $S(t_1, t_2)=\sigma_e^2 I(t_1=t_2)$.  %While easy to implement, some weaknesses of these multi-step approaches are that choice of regularization parameter is challenging, and they have the potential to be inefficient since replication and regularization are done separately, not able to inform each other. 
 Wu \& Chiang (2000) fit a general linear functional response regression model, assuming iid errors, no random effect functions, and using kernel smoothing to regularize the functional coefficients, and Chiang et al. (2001) proposed smoothing spline regularization for the same model.

Reiss et al. (2010) studied Ramsay and Silverman's approach to general functional response regression for curves sampled on common, fine grid, with basis representations (splines) for the observed curves and functional coefficients and regularization done by roughness penalties, while ignoring within-function correlation.  They called this method ordinary least squares (OLS), and compared with an approach where the within-function covariance was estimated in the basis space in an unstructured fashion, and accounted for in estimation by WLS.   They also introduced automatic smoothing parameter estimation methods and inferential techniques for this model.  Shi et al. (2007) specified a general functional response regression model using Gaussian processes to capture the within-function covariance structure $S$.  They used a multi-step approach to fit the model, representing each observed function using a $K_y$-dimensional B-spline decomposition, performing a regression for each B-spline coefficient, and then fitting a Gaussian process to the empirical curve-to-curve deviations.  Krafty et al. (2008) worked with model (\ref{eq:FRR1}) %varying coefficient models with general fixed effects structure, 
using PC decompositions to capture the within-function covariance.  Functional coefficients were modeled by smoothing splines with roughness penalties and estimated by IRLS, with inference done by point-wise confidence intervals.  %and estimated through penalized generalized least squares, taking the within-function covariance into account in their estimation and updating the functional coefficients and estimating the within-function covariance using iteratively reweighted least squares (IRLS).   Inference is done via pointwise confidence intervals.  
They mentioned that smoothing spline estimates of functional coefficients from correlated data are consistent but inefficient if the within-function correlation is ignored (Lin et al. 2004).% (Welsch et al. 2002; Lin et al. 2004).  

% Most functional response regression methods have assumed linear relationships between predictors and functional responses.  One exception is Chiou et al. (2003), who specified a semiparametric method for functional response regression model intended for functions sampled on a common grid.  The multi-step fitting approach involves first estimating the mean function pointwise, smoothing it, estimating the between-function covariance matrix on the grid, estimating the eigenvectors and then smoothing them.  Their regression involves calculation of the mean of the curve-level PC scores conditional on a set of $p$ predictors using a single-index model, with model fitting done using an iterative quadratic linear unbiased estimation approach.  Their model has white noise errors and no curve-level random effect functions, so $S=\sigma^2_e I$.  Chiou \& M\"uller (2004) extended this approach to multiple-index models, and Chen et al. (2011) improved the approach for estimating the link function.

 \textbf{Modeling Correlated Functional Data and Functional Mixed Effects Models:} 
  All of the aforementioned methods assumed an independent sample of functions, and cannot capture between-function correlation induced by the experimental design.  This correlation is present, for example, in multi-level functional models for which multiple functions are observed for each subject or within clusters such as the \textit{DTI data set} in Section \ref{sec:data}, in longitudinally observed functional data where functions are observed serially in time or some other variable like pressure in the \textit{glaucoma data set} of Section \ref{sec:data}, or in spatially correlated functional data where functions are observed for each unit on a spatial grid.  It is important to take the induced correlation between functions into account in estimation, to ensure observed functions are properly weighted in the regression through WLS, and inference, to ensure that standard errors, point-wise bands, or joint bands properly account for all sources of variability in the data.  
  
 One way to capture this correlation is to specify a between-function covariance structure $R$ in the curve-to-curve deviations in model (\ref{eq:FRR1}) that is separable from the within-function covariance $S$, for example assuming the total covariance is given by a tensor product $R \otimes S$, implying COV$\{E_i(t_1),E_{i'}(t_2)\}=R_{ii'}S(t_1,t_2)$.  This strategy has been used in many time-space models in the spatial modeling literature, and is assumed in the context of spatially correlated functional data by the separable version of the functional CAR model introduced by Zhang et al. (2014) which assumes a block conditional autoregressive (CAR) structure for $R$. % Morris \& Carroll 2006.  %or row tensor product ($R \odot S$=\{$R \otimes I_T+I_N \otimes S\}$, Scheipl et al. 2014).  
 
 Alternatively, correlation between functions can be induced by adding random effect functions to the general linear functional response regression model (\ref{eq:FRR1}) to form \textit{functional mixed effect models} (FMM).  For example, consider the following FMM that was introduced in this form by Morris and Carroll (2006):
  \begin{equation}
Y_i(t_j)=\sum_{a=1}^p X_{ia} B_a(t_j) + \sum_{h=1}^H \sum_{l=1}^{M_h} Z_{ihl} U_{hl}(t_j) + E_i(t_j),\label{eq:FMM1}
 \end{equation}
 where $H$ is the number of levels of random effects, $Z_{ihl}$ are random effect covariates at level $h$ with corresponding random effect functions $U_{hl}(t)$.  Suppose the $U_{hl}(t)$ are iid (over $l$) mean zero Gaussians with covariance $Q_h(t_1, t_2)$ representing the within-function covariance structure at random effect level $h$.  Although conditionally independent, these random effect functions can induce correlation between the functions through the structure of their design matrices.  
 
 To illustrate, let $\mathbf{Y}(t)=(Y_1(t), \ldots Y_N(t))'$, $\mathbf{B}(t)=(B_1(t), \ldots, B_p(t))'$, and $\mathbf{Z}_h$ be an $N \times M_h$ matrix with element $(i,l)$ given by $Z_{ihl}$.  For a given $t$, the marginal $N \times N$ covariance matrix for $\{\mathbf{Y}(t)|\mathbf{B}(t)\}$ is given by $\sum_h \mathbf{Z}_h \mathbf{Z}_h' Q_h(t,t) + S(t,t) I_N$, which introduces between-function correlation through the random effect design matrices $\mathbf{Z}_h$, with each component $\mathbf{Z}_h \mathbf{Z}_h' Q_h(t,t)$ capturing a different source of between-function variability.  For example, a design matrix $Z_{ihl}=1$ only if curve $i$ is from cluster $l$ induces a compound symmetry correlation structure among all curves within the cluster because of the shared cluster-specific random effect function $U_{hl}(t)$.  For longitudinally observed functional data, if $d_i$ represents the longitudinal time point at which curve $Y_i(t)$ is observed, then by defining $Z_{ih'l}=d_i$ if from subject $l$, 0 otherwise, and with $U_{h'l}$ representing the random slope for subject $l$, this structure will capture serial correlation across the longitudinally observed curves.   Notes: [1] These between-function correlations cannot be induced by curve-level random effects, i.e. if all random effect functions are indexed by curve, $U_{ihl}(t)$, which is an important distinction between model (\ref{eq:FMM1}) and the type of functional mixed effect model introduced in Guo (2002).   [2] When the structure of $Q_h$ and $S$ allow heteroscedasticity across $t$ (i.e. $Q_h(t,t)$ and $S(t,t)$ vary over $t$), then the level of between-function correlation is allowed to vary over the functional index $t$. [3] When $Q_h$ and $S$ allow within-function correlation (i.e. are not diagonal), then the $N \times N$ marginal cross-covariance of $\mathbf{Y}(t_1)$ and $\mathbf{Y}(t_2)$ is given by $\sum_h \mathbf{Z}_h \mathbf{Z}_h' Q_h(t_1, t_2) + S(t_1, t_2) I_N$.
 
 Additional between-function correlation can be induced in functional mixed models by specifying a separate between-function covariance $P_h$ across the $M_h$ random effect functions at level $h$, and then using tensor products ($P_h \otimes Q_h(t_1, t_2)$, Morris and Carroll 2006) or row tensor products ($P_h \odot Q_h (t_1, t_2) = \{P_h \otimes I(t_1=t_2) + I_N \otimes Q_h(t_1, t_2)\}$, Scheipl et al. 2014) for the total covariance of $U_{hl}(t_1)$ and $U_{hl'}(t_2)$.  % that is if curve-level random effects  %If curves are sampled longitudinally within clusters, with curve $i$ sampled at time $s_i$, then one could capture the longitudinal correlation across curves by adding a level of random slope functions for each cluster, in which case $Z_{ihl}=s_i$ if from cluster $l$, 0 otherwise.
 %Further, the marginal cross-covariance COV\{\mathbf{Y}(t_1),\mathbf{Y}(t_2)|\mathbf{B}(t_1),\mathbf{B}(t_2)\}= $\sum_h \mathbf{Z}_h \mathbf{Z}_h' Q_h(t_1, t_2) + S(t_1, t_2) I_N$, showing how the within-function covariance assumptions come into play.
%An alternative approach to capturing between-function corrrelatio the random effect functions or residual errors can be made dependent across curves, specifying separate between-function covariance structures $P_h$ across the $M_h$ random effect functions at level $h$ and across the residual error functions ($R$), using tensor products (Morris \& Carroll 2006) or tensor sums (Scheipl et al. 2014) to combine the between-function ($P_h$ or $R$) and within-function ($Q_h$ or $S$) covariance structures.
 
\newpage

\textbf{Different types of Random Effects in Functional Response Regression Models:} To clarify an important issue in this context, random effect model components are used in various ways in functional regression, and not all of them capture between-function correlation.  First, in single function spline smoothing, random effects on the spline coefficients are used as a mechanism of penalization to induce regularization within the function, which could be called \textit{random effects for penalization}.  These are called \textit{new style random effects} by Hodges (2013, Chapter 13), and most familiar to spline smoothers.  Second, in functional response regression the \textit{curve-level random effect functions} defined in model (\ref{eq:FRR1})  capture within-function correlation in the curve-to-curve deviations, but again the purpose is strictly regularization, as these curve-level random effects, if independent, cannot induce correlations among the functions.  %This use of random effects is most familiar with spatial modelers using Gaussian Processes, and are commonly used in the functional growth curve and FANOVA methods mentioned above. 
Third, random effect functions can be defined for units of observations such as clusters or subjects for which multiple curves are observed.  These could be called \textit{multi-level random effect functions}, and serve multiple purposes.  %They are a combination of both the \textit{old-style} and \textit{new-style} random effects mentioned by Hodges (2013, Chapter 13), and serve multiple purposes.  
They capture multiple sources of between-function variability in the data and induce between-function correlation, so are involved in \textit{replication}, determining the manner in which information is combined across functions in performing functional response regression.  They are also involved in \textit{regularization}, since within-function covariance assumptions for each level of random effect functions define how strength is borrowed within a function which impacts estimation and inference of the functional coefficients.   These are the type of random effect functions in model (\ref{eq:FMM1}).%, and have to date not been used by many researchers.  %These are the type we are referring to in functional mixed models.

In classic linear mixed effect models, random effects are integrated out and fixed effects are estimated using marginal likelihood, leading to a WLS-like fitting procedure that  re-weights the observations according to the covariance structure induced by the random effects.  As a result, inference on the fixed effects integrates over the populations sampled at the random effect levels, allowing inference to be made on that reference population.  For some methods involving functional mixed models, the random effect functions are not integrated out of the model but instead conditioned upon in an additive multi-step fitting procedure.  In these cases, there is no integration over these sources of variability in estimation and inference of the fixed effect functions.  Thus, these \textit{conditional random effect functions} behave much like fixed effect functions, providing smoothed estimates for specific individuals but not allowing inference on the the fixed functional regression coefficients to cover the reference population from which they were drawn.% from which they were drawn, which has inferential implications for the fixed functional coefficients.%, and when this is done the model has no WLS-type reweighing or inferential accounting of between-function correlation inherent to proper mixed effect modeling, so these effects behave like .  

\textbf{Papers involving Functional Mixed Effect Models:}  Brumback \& Rice (1998) introduced the first published approach for modeling multi-level functional data, a smoothing-spline based method for nested or crossed curves.  %applied to menstrual cycle data where progesterone curves for menstrual cycles are nested within subjects, and subjects within groups.  
Their model contains mean functions at the group level, plus deviation curves at the subject and curve-within-subject levels, all represented by smoothing splines using the Demmler-Reinsch parameterization plus white noise errors.  They split the random functions at each level (group, subject, curve) into the sum of fixed intercepts and slopes and random iid Demmler-Reinsch spline coefficients.  The between-function correlation at the different levels is captured by these random spline coefficients, but the the form of the subject-level ($Q$) and cycle-level ($S$) within-function covariance matrices are restricted, fully determined by the spline basis and roughness penalty up to a single variance component, the smoothing parameter.  %Fitting is done by EM.

Guo (2002) first introduced the term \textit{functional mixed effects models} for the model presented in that paper which includes arbitrary numbers of fixed effect functions and $q$ curve-level random effect functions, all modeled as smoothing splines, and iid errors.   Since this model includes only curve-level random effect functions, it cannot capture between-function correlation or accommodate multi-level models like the FMM (\ref{eq:FMM1}) presented above.  Splitting the random effect functions into linear and spline components, they modeled intercepts and slopes as random effects, and so their within-function covariance $S$ is determined by the sum of a parametric part induced by the covariance of the random intercept and slope, a spline part whose form is fixed by the spline basis and penalty up to the scalar smoothing parameters, plus white noise.  They fit their model using either PROC MIXED or a faster Kalman-filter based approach, and described pointwise confidence bands and functional testing based on RLRTs.  Qin \& Guo (2006) extended this approach to induce periodic assumptions on the fixed and random effect smoothing splines.
 
Morris et al. (2003) presented a Bayesian, wavelet-based method for nested functional data with three hierarchical levels: group, subject within group, and curve within subject, with fixed mean functions for each group and random effect functions at the subject and curve levels.  They assumed diagonal covariance matrices for the wavelet coefficients of the random effect functions at the subject and curve levels, $Q^*$ and $S^*$ respectively.  The variance components were indexed by wavelet scale but not location, constraining the form of the within-function covariances $Q$ and $S$ to  stationary-type structures.   They presented a fully Bayesian model for estimation and inference.  Fixed effect functions were regularized using SSVS priors on the wavelet coefficients, and their estimation accounts for the between-function correlation induced by the random effect functions in a WLS-like fashion.

Morris \& Carroll (2006) extended this approach to the general FMM (\ref{eq:FMM1}) with general design matrix $X$, multiple levels of random effect functions, plus curve-to-curve deviations, in a method intended for functional data sampled on a common fine grid.  Using a \textit{basis transform modeling approach}, they computed $T^*\approx T$ wavelet coefficients for each observed function, thus projecting the data into the wavelet space, and fit a version of the FMM in the wavelet space, with diagonal but heteroscedastic assumptions for $Q_h^*$ and $S^*$ across wavelet coefficients.  Unlike Morris, et al. (2003) who only let variance components vary across wavelet scale, in this work they allowed each of the $T^*$ wavelet coefficients to have its own variance component.  While limiting the form of the within-function covariances $Q_h$ and $S$ relative to unstructured covariances (which would not be feasible to fit in most settings), this structure is flexible enough to capture important nonstationarities such as heteroscedasticity and different degrees of within-function autocorrelation across $t$.  This greatly speeds calculations, making the procedure $O(T^*)$ and allowing it to scale up to perform fully Bayesian analysis for enormously sized functional data sets such as mass spectrometry (Morris et al. 2008), quantitative image data (Morris et al. 2011; Fazio et al. 2013), or whole-genome analyses (Lee \& Morris, 2014).  

Fixed effect functions are adaptively regularized through sparsity priors in the wavelet space, and random effect functions are adaptively regularized through the wavelet-space Gaussian distributions with coefficient-specific variances that behave like scale mixtures and thus have adaptive shrinkage properties (Morris \& Carroll 2006).  The random effect function and curve-to-curve deviation covariance structures involve respective tensor products of between- and within-function covariances $P_h \otimes Q_h$ and $R \otimes S$, respectively, which allows highly structured between-function covariances, e.g. for spatially correlated functional data (Zhang et al. 2014).  Even when conditionally independent random effect functions are used ($P_h=I_{M_h}, R=I_N$),  the multiple levels of functional random effects with general design matrices enable great flexibility in capturing between-function correlation, for example modeling nested structures through random functional intercepts or modeling longitudinally observed functions through random functional linear coefficients that are parametric in time but nonparametric in the functional index (Lee et al. 2014).  Further, by adding random effect levels with design matrices consisting of Demmler-Reinsch spline bases constructed for some continuous predictor $X_i$,  the model can also incorporate nonparametric functional additive terms $f(X_i,t)$ in place of functional linear terms $X_i B(t)$ (Lee et al. 2014). %serve as adaptive shrinkage penalties.  %This highlights the interplay of replication and regularization in random effect regularization.  The random effect Gaussian VC vary by wavelet coefficient, so coefficients with a large degree of variability are shrunken less, while those with less variability are shrunken more, leading to nonlinear shrinkage and adaptive regularization.    
The fully Bayesian modeling approach integrates over all stochastic levels, yielding WLS-like fixed effect function estimates and inference that account for the various sources of between- and within-function variability captured by the model.    Joint credible bands can be computed, from which functional hypotheses can be tested (Meyer, et al. 2013),  and pointwise posterior probabilities, interpretable as local false discovery rates (FDR), can be computed to flag regions of the fixed effect curves as significant while controlling global FDR (Morris et al. 2008).  %Their method is intended for functions sampled on common fine grids, not sparsely observed data on different grids, but can scale up to enormous data sizes for fully Bayesian analysis of proteomics, whole genome assays, and high dimensional neuroimage data sets.  Lee and Morris (2014) applied the WFMM to whole genome methylation data, and demonstrated by simulation that accounting for the within-genome correlation through wavelet bases leads to more power to detect differentially methylated regions than methods that model genome positions independently.  

Morris et al. (2006) extended this method to deal with partially missing functions, i.e. with missing grid points for some of the functions, and Morris et al. (2011) extended this framework to 2d or 3d image data.  %They also introduced a joint compression approach to select a common set of $T^*$ basis functions needed to account for a minimum percent total energy for all functions that reduces calculations from O($T$) to O($T^*$), with $T^*\ll T$ for sparse bases like wavelets or PCs.  
While the method was introduced using wavelet bases, subsequent work (Morris et al. 2011; Meyer et al. 2013; Zhang et al. 2014; Lee et al. 2014) has described how the method can be applied using other basis functions such as Fourier bases, PC, and splines while using the same basis transform modeling strategy.  The forms of the within-function covariances at each level are determined by the choice of basis functions up to a set of $T^*$ coefficient-specific variance components.  Regularization is done by either truncation of bases using a joint basis truncation strategy introduced in Morris, et al. (2011), and/or through roughness or sparsity penalties in the basis space, both of which can be induced as special cases of a mixture prior consisting of a spike at 0 and a Gaussian scale mixture.  Zhu et al. (2011) developed a robust version of this Bayesian FMM that uses heavier-tailed distributions for the likelihood and random effect distributions to produce functional coefficient estimates and inference that are insensitive to outliers.  As discussed below, this framework can also be used to perform general function-on-function regression (Meyer et al. 2013) and analyze spatially correlated functional data (Zhang, et al. 2014).

Antoniadis \& Sapatinas (2007) presented a FMM with general fixed effect functions, curve-level random effect functions, and iid residual errors, using wavelet bases to represent the functions.  Unlike Morris \& Carroll (2006) who focus primarily on estimation and point wise inference on the fixed effects, their main focus is on functional inference.   %Given the random effects are at the curve level, their model cannot capture between-function correlation or handle multi-level models, although they note that the model can be adapted to do so.  %However, they use the same notation of Guo (2002) regarding q random effects per curve i.  
%Like Guo (2002), 
%Like Guo (2002), their functional mixed effect model has only curve-level random effect functions, so cannot capture between-function correlations.  
Their random effect curves are assumed iid with within-function covariance matrix $S$ completely determined by the assumed Besov space up to a single scalar variance component.  This along with the iid error assumptions limit its flexibility, but allows a simple functional hypothesis test for the significance of the random effects.  %the testing of random effects using a simple hypothesis test.  
They do not apply any regularization to the fixed effect functions.
%Like Morris et al. (2003) and MC (2006), they model in the wavelet space, but without any regularization in the wavelet space.  Unlike Morris and Carroll (2006) who focus on estimation and pointwise inference, they focus their efforts on construction of functional inference.

Berhane \& Molitor (2008) introduced a complex, multi-level functional mixed model for growth curves of individuals nested within clusters.  Their model included fixed effect functions for overall intercepts and covariate effects, plus functional random effect intercepts and slopes for covariate effects at both the cluster and random effect levels, with iid errors.  The fixed and random effect functions were parameterized using natural cubic B-splines, with regularization done by truncation inherent in the knot selection.  The used a two-step method to fit this model.%They acknowledge this complex model is challenging to fit, so introduce a Bayesian two-stage approach that involves fitting separate models for each cluster, and then a second stage model that updates the population-level parameters while integrating over the uncertainty from the first-stage model.  
%Scarpa \& Dunson (2009) presented a Bayesian method for multi-level functional data involving multiple ovulation cycles for each of a number of women.  They represent the fixed and random effect functions at the various levels using a parametric piecewise linear model motivated by the application, but with a nonparametric contaminant component modeled through a Dirichlet process.  

Aston et al. (2010) presented a PC-based FMM for functions sampled on common grids.  After subtracting a pointwise (unsmoothed) mean function, they estimated the eigenfunctions based on the empirical covariance matrix, and then computed the PC scores for each curve, truncating at $T^*$ coefficients, effectively projecting the data into the space spanned by the overall PCs.  They next fit independent LMMs for each PC score, regularizing through sparsity by performing stepwise variable selection across the PC dimensions for each fixed effect function.  Their model included a single level of random effect functions at the curve level. %although it is clear their approach is extendable to multiple random effect levels to be used with multi-level functional data.

Chen \& Wang (2011) presented a spline-based FMM for sparse functional data with general functional fixed effect structures, curve-level random effects, and residual errors with a parametric covariance (e.g. AR(1)) that is heteroscedastic across $t$.   They represented the functional coefficients, random effects, and the log of residual variances (varying over $t$) using spline bases with roughness penalties for regularization.  They nonparametrically estimated the random effect functions and an unstructured covariance for the random effect spline coefficients $S^*$ using an EM algorithm, and accounted for the within-function covariance by updating the fixed effects using a WLS-like approach.  With only curve-level random effects, their model cannot accommodate between-function correlation.

Greven et al. (2012) presented a model for longitudinally observed functional data $Y_{ij}(d_{ij}, t)$, with $d$ representing the longitudinal index and $t$ the functional index.  Their model includes an overall mean function $\nu(d_{ij}, t)$ and three levels of random effect functions: random slope and random intercept functions at the subject level, and a curve-level random effect function.  Their primary focus is simply estimation of the ML-fPC at these 3 levels and their accompanying scores, which they do after subtracting off a 2d-kernel smoothed estimate of the mean function that does not take the between- or within-function covariances into account.  Goldsmith \& Kitago (2013) introduced a Bayesian method for general functional response regression for nested data.  Their model has general fixed effect functions, represented by B-splines regularized by a roughness penalty, plus subject-level random effect functions and curve-to-curve residual error deviations.  Their subject-level random effect functions are represented by penalized splines, with their effective within-function covariance structure $Q$ determined up to a scalar constant by the spline and penalty choice.  The within-function covariance $S$ for the curve-to-curve deviations is left unstructured and given an inverse Wishart prior.  Estimation is done using Variational Bayes.

\textbf{Spatially Correlated functional data:}  Some papers have dealt specifically with functional response regression in settings where the observed functions are correlated according to some spatial grid.  Within a nested multi-level functional model with correlated functions within subjects and subjects within groups, Baladandayuthapani et al. (2008) introduced a Bayesian method using truncated power series spline bases to represent the functions at each level of the hierarchy, and capturing the correlation between curves using a Matern structure on the spline coefficients.  Their within-function covariances at each hierarchical level $Q_h$ consisted of the sum of random polynomial terms plus a spline part whose structure is determined by the choice of basis and penalty up to a scalar smoothing parameter.   Staicu et al. (2010) presented an alternative approach to this model, using ML-fPC bases instead of splines at each level of the hierarchy, polynomials for the group mean functions, and estimating the spatial variogram using method-of-moments.   The ML-fPCs yield flexibility in the within-function covariance assumptions at each level, and their use of WLS to update the group mean functions integrates over the different levels of variability in the model.  Their fitting was done using a multi-step approach that involves sequential estimation of the variogram, ML-fPC bases, group mean functions, and the residual error variances, yielding point estimates of each quantity.  Zhou et al. (2010) also used ML-fPC at each level of the hierarchy, but instead of parametric mean functions used polynomial splines to nonparametrically represent the mean functions and ML-fPC, with regularization by roughness penalties.  They captured the between-function spatial correlation at the curve level by specifying a stationary covariance structure among the curve-level ML-fPC scores.  Unlike Baladandayuthapani et al. (2008) and Staicu et al. (2010), with these assumptions their model allows non-separable spatial correlations when projecting the correlation from the fPC space back to the function space.  In the context of areal functional data, Zhang et al. (2014) introduced a functional CAR model.  The separable version of their model was previously mentioned above, and they also introduced a more flexible non-separable version that projects the functions to a chosen basis space and fits conditional autoregressive (CAR) models in the basis space with separate CAR parameters for each of $T^*$ basis coefficients.  This construction leads to a CAR model between the functions that allows the spatial correlation to vary across $t$, yet borrows strength across $t$ in estimating the correlation as determined by the chosen basis functions.  Embedded within the general FMM framework in (\ref{eq:FMM1}), this allows functionally varying spatial correlation to be accounted for in general functional response regression.

\textbf{Semiparametric Functional Mixed Models:} Most of the existing functional response regression methods involve fixed effect functions that are nonparametric in $t$, but linear in $x$.  There is recent work extending the idea of generalized additive models (GAM, Hastie \& Tibshirani 1986) to functional response regression leading to a new class of semiparametric models for functional response regression that include terms than be either parametric or nonparametric in $x$.  In the last chapter of Wood (2006), additive mixed models (AMM) are described as an extension of Laird and Ware (1982) with generalized additive fixed effects and parametric random effects.  If parametric linear forms are assumed in $t$ for the fixed and random effects, then AMM can be used to fit parametric functional additive mixed models to estimate terms such as $f(x,t)$ representing an effect of $X$ on $Y(t)$ that is nonparametric in $x$ but parametric in $t$.  The models are fit using $mgcv$ in R, which utilizes $lme$ to fit the LMMs, and it is mentioned that computational problems can arise when fitting very complex models. %which he mentions is worked hard mentions that these models work $lme$ hard, and can have numerical problems when fitting complex models.

Scheipl et al. (2014) described a generalization of AMM to the functional response setting that allows $f(x,t)$ to be nonparametric in $x$ and $t$.  They introduced an elegant notation for specifying design matrices for an impressively general array of fixed effect function structures, including intercepts, linear, or smooth nonparametric effects for scalar predictors with constant or functional (in $t$) effects, plus linear or smooth nonparametric effects for functional predictors.   Their notation subsumes the fixed effect structures used in current literature.  Their construction uses row tensor products of between- and within-function design matrices, each with their own roughness penalty that consists of a scalar smoothing parameter and specified penalty matrix.  It is primarily designed to be used with penalized splines for the smooth effects and functional coefficients, each with their own separate roughness penalties for regularization.    Their construction assumes iid errors but also allows multiple levels of random effect functions, whose covariance structure is a row tensor product (see definition above) and is completely determined by the choice of basis functions and specified between- and within-function penalties (serving as covariance matrices) except for two scalar variance components that are the between- and within-function smoothing parameters.  %Penalized splines are used for the random effect functions, although if the model is limited to a single level of random intercepts, principal component bases can be used with the eigenvectors and scores estimated up front and placed in the design matrices.  
Given $N$ functions each sampled on a grid of size $T$, they simplify the entire model down to a large AMM or LMM of size $NT$ that is fit using $mgcv$  from Wood(2011), from which pointwise intervals and RLRT tests can be computed.  The use of a single large model limits the size of data sets that it can feasibly accommodate, so %compared with Morris and Carroll (2006) and Morris, et al. (2014) 
as they mention it does not scale up well to functional data sets with enormous $T$, e.g. the \textit{mass spectrometry data set} in Section \ref{sec:data}, but works well for sparsely sampled functions and can accommodate unequal sampling grids between functions.  %One advantage of their approach is that they can accommodate varying sampling grids across functions.  
%The paper does not state whether their computational approach integrates over the random effect functions when performing estimation and inference of fixed effect functions using a WLS-like approach or effectively estimates them conditionally in an additive fashion.  %It is not clear in the paper whether their approach effectively integrates over the random effect functions when computing estimates and inference for the fixed effect functions, or if inference is based only on the iid residual error variance.  

As described above, Lee et al. (2014) showed how the Bayesian FMM of Morris and Carroll (2006) and following can accommodate additive coefficients $f(x, t)$ that are nonparametric in $x$ and $t$ by adding a level of random effect functions with corresponding design matrix defined based on Demmler-Reinsch spline bases on $x$, with the variance component regulating the smoothing in $x$ allowed to vary by basis coefficient $k$.  By allowing the smoothing parameters to vary across the response basis coefficients, their approach allows the degree of smoothness in $x$ to vary and borrow strength over $t$.  Their Bayesian modeling approach allows full joint or point-wise inference to be done on the coefficient surface or linear functionals, e.g. derivatives.  This method is designed for functions sampled on a common fine grid.

\textbf{Spectral domain FDA of time series.}  A number of papers apply FDA approaches to time series using spectral domain modeling, which effectively involves projecting the data to the Fourier domain and modeling there.   Diggle \& Wassel (2007) used Fourier domain modeling involving the log-spectrum for modeling of dense growth curves.  Guo (2003) performed smoothing-spline ANOVA on log periodograms, with separate periodograms for partitions of time to allow locally stationary processes.  They used tensor product spline bases to model the coefficients with roughness penalties for regularization, and accommodated general fixed effect specifications.  Qin et al. (2009) applied a state-space modeling approach to scale this approach up to large data sets, and applied it to EEG data.  Freyermuth  et al. (2010) estimated growth curves in the spectral domain by representing the overall mean and curve-specific deviations of log-spectra using tree-structured wavelets, with adaptive regularization done by truncating to a limited number of wavelet coefficients, keeping the same set of wavelet coefficients for each curve and the overall mean.  Krafty et al. (2011) fit a functional mixed effects model in the spectral domain for stationary time series, with general fixed effect functions and handling two-level nested designs in the random effects, with $n_j$ time series from each of $N$ subjects, and random effect spectra for each of the $N$ subjects.  They have $q$ random effects per subject, each with their own covariance.  The covariance was estimated by first estimating all of the subject-level random effect functions as smoothing splines, with a common smoothing parameter across subjects, % inducing a covariance matrix that is fixed by the spline basis up to that constant smoothing parameter, 
and then estimating the covariance nonparametrically from these smooth estimates. % It is questionable whether these are proper random effects since they are not estimated as BLUPs but just the same as fixed effect smoothing splines, except that they share a common smoothing parameter.  
Inference was done by pointwise confidence bands.  Their approach can be extended to locally stationary time series by partitioning time and using the tensor product approach of Guo (2003).  Martinez et al. (2013) applied the Bayesian wavelet-based image FMM of Morris et al. (2011) to spectrogram images summarizing time-varying spectral content, allowing functional response regression of nonstationary time series on multiple predictors of various types, while accounting for potential correlation among the time series induced by the experimental design through random effect functions.

\textbf{Functional Discriminant Analysis using Functional Response Regression:}  Although functional response regression models treat the function $Y_i(t)$ as the response and any class indicator $X$ as the predictor, after fitting functional response regression models one can perform functional discriminant analysis to classify subjects and curves based on their functional data, in which case the functional response regression model serves to characterize a parametric joint density for the functions within each group.  This has been done in various modeling contexts.  Do \& Kirk (1999) estimated the mean curve through penalized splines with roughness penalties, and used Fourier domain-smoothed PCs to model the within-function covariance matrices for each group.  Hall  et al. (2001) performed a PC decomposition using the combined data set including all groups, and constructed discriminant functions on the PC Scores.  James \& Hastie (2001) used spline representations for the mean functions, and PCs for the errors to construct their discriminant function.  Li \& Yu (2008) introduced a method %\textit{functional segment discriminant analysis }(FSDA), 
combining linear discriminant analysis and support vector machines.  Bayesian approaches for functional discriminant analysis have the added benefit of being based on predictive distributions, integrating over the uncertainty of the model parameters rather than simply conditioning on estimates of them.  Zhu et al. (2012) introduced Bayesian methods to perform functional discriminant analysis in the FMM framework of Morris \& Carroll (2006).  This allows the model to adjust for other covariates affecting the functions and enables classification based on a set of correlated functions.  Methods were described for both the Gaussian and robust (Zhu et al. 2011) versions of the FMM, and for use with 1d functions or higher dimensional functions like images, using wavelets or any other chosen basis modeled using a basis transform modeling approach.  %They mention their approach can be used with any functional response regression model.  
Stingo et al. (2012) presented a Bayesian, wavelet-based method for functional discriminant analysis.  They modeled in the wavelet space, using an Inverse Wishart prior on the covariance for each group to allow more general covariance structures than approaches assuming independence in the wavelet space.   %While of some importance for estimating functional response regression coefficients, clearly the flexible and realistic handling of within-function correlation is crucial in the context of functional discrimination.
 %%% Mnetion BAyesian methods use predictive rather than conditional framework for discrimination.

\textbf{Summary of Functional Response Regression Models:}  The literature and software for functional response regression is much less developed than that for functional predictor regression.  A vast majority of the published papers simply involve estimation of a mean of a sample of functions, which has been called \textit{growth curve analysis} in classic literature.  These methods differ in their approach for smoothing the mean function, with different choices of basis functions or regularization approaches, and also in their approach for modeling the correlation over $t$ in the curve-to-curve deviations, if they even model such correlation.  Of the methods for general functional response regression, most cannot model correlation between curves, so are only suitable for independently sampled functions.  Again, methods differ in choice of basis functions and assumptions made on the curve-to-curve deviations. Methods that assume independent and identically distributed residual errors and no curve-to-curve deviations to capture the within-function correlation are not realistic for the FDA setting, since the nature of the functional data suggests there should be correlation across $t$ and often heteroscedasticity in the curve-to-curve deviations.  These may yield reasonable point estimates but the inference may be questionable because of the incorrect independence and homoscedasticity assumptions.   Many of the spline-based functional response regression models accommodate within-function correlation in the curve-to-curve deviations, but through the choice of basis function and penalty make assumptions on the within-function covariance that fixes its structure up to a constant, the smoothing parameter.  These types of simplifying assumptions are necessary in single-function spline smoothing, but not in the functional data analysis setting, since the replicate functions allow the estimation of more rich structures in the covariance of the curve-to-curve deviations.  Other methods accommodate more flexibility in these matrices, either modeling them principal component decompositions, unstructured covariance matrices, or with parsimonious but still somewhat flexible assumptions induced by assuming independence in a chosen basis space.  The covariance assumptions in the curve-to-curve deviations may be important, as they can affect the inference on the functional regression coefficients, and also strongly come into play when performing functional discriminant analysis.  

There is a small but growing literature on methods that can model correlated functions in the functional response regression setting.  This includes models for spatially correlated functional data, longitudinally observed functional data, as well as general functional mixed models that can include various levels and types of random effect functions that can induce correlation both between and within functions, and take these correlations into account in constructing point estimates and inference for the functional coefficients.  Most of the methods are intended for functional data consisting of smooth, simple functions sampled on a coarse grid, and make modeling assumptions that are not flexible enough and/or computationally scalable to settings with complex functional data sampled on fine grids, for example spiky functions like mass spectrometry data (Morris, et al. 2008) or quantitative image data like functional magnetic resonance imaging (Morris, et al. 2011).  These complex, high-dimensional data often have richly structured with-curve covariance matrices, yet in many of these settings their complexity along with the fact that $T \gg N$ raise questions about whether principal components can be well-enough estimated to be reliably used.

 \textbf{Available Software for Functional Response Regression:}  Only a few of the discussed functional response regression methods have available software.  Standalone executable software for the Bayesian FMM methods introduced by Morris and Carroll (2006) are available at \url{https://biostatistics.mdanderson.org/SoftwareDownload/SingleSoftware.aspx?Software_Id=70} for Windows, Linux, or OS-X.  With properly defined design matrices $\bX$ and $\mathbf{Z}_h$, this standalone can be directly used to implement the MCMC in the methods presented in Morris et al. (2006), Morris, et al. (2008), for 2d or 3d image responses in Morris et al. (2011), function-on-function regression in Meyer, et al. (2013), and the semiparametric functional mixed models and longitudinally observed functional modeling approaches in Lee, et al. (2014).  The current version uses wavelet bases, although one can fit Bayesian FMMs using other basis functions if they compute the transforms and inverse transforms outside the package and input the data matrix in the basis space.  %, and the next version will be able to internally handle other basis functions including principal components and pre-specified basis transform matrices.  
  Broken down into preprocessing, MCMC, and post processing modules, the executable is parallelizable, allowing it to be used with enormous data sets.  As described in Herrick and Morris (2006), this method is at worst linear in the number of observations per function $T$, and even faster when joint compression (Morris, et al. 2011) is used to perform truncation on the basis coefficients.  The method automatically produces a set of posterior summary statistics (including means, standard errors, point wise and joint credible bands, and point wise posterior probabilities of pre-specified effect sizes)  for each fixed effect function and pre-specified contrast of fixed-effect functions for each $t$ or specified linear combinations across the $t$, plus saves the posterior samples so any other Bayesian estimation, inference, prediction, or diagnostics can be computed.  %The work is ongoing and the package will be updated to include other methods in this vein of Bayesian FMM work (Zhu, et al. 2011; Zhu, et al. 2012; Martinez, et al. 2013; Zhang, et al. 2014) and an R wrapper for it, as well.
 
The smoothing spline functional mixed model of Guo (2002) can be applied using a SAS macro at \url{http://www.jstatsoft.org/v43/c01}.  The method for spatially correlated hierarchical functional data of Baladandayuthapani, et al. (2008) has software available at \url{http://works.bepress.com/veera/5}.   The fast function-on-scalar regression methods of Reiss, et al. (2010) is implemented in the R package $fosr$ in the $refund$ package, along with an implementation of Fan and Zhang (2000)'s two-step approach in the $fos2r$ package.  

Wood (2006) has implemented his generalized additive model approaches using $mgcv$, which includes a package $gam$ to fit generalized additive models, which uses $lme$ in R to fit the linear mixed models.  The R package $pffr$ in $refund$ contains a wrapper which sets up the models of Scheipl, et al (2014) in a way so they can all be fit using Wood's $mgcv$ package in R.
 
\section{Function-on-Function Regression} \label{sec:FFR}

There has been comparatively little work on function-on-function regression done to date.  Ramsay \& Dalzell (1991) introduced the functional linear model, which is a function-on-function regression model with unconstrained surface coefficient $B(s,t)$:
\begin{eqnarray}
Y_i(t)=B_0(t) + \int X_i(s) B(s,t) ds + E_i(t). \label{eq:FonF}
\end{eqnarray}
This setting combines together the issues present in functional predictor and functional response regression settings, including regularization of the predictor function for denoising, reducing collinearity, and dealing with variable grids across functions, regularization of the coefficient surface in both dimensions in order to obtain more interpretable and efficient estimators, and regularization in terms of modeling the within-function correlation in the residual errors and taking this into account in estimation through WLS-type approaches and in estimation.  Additionally, when the response curves are correlated, random effect functions or correlated errors may be necessary to take between-function correlation into account. 

Ramsay \& Dalzell (1991) fit model (\ref{eq:FonF}) using piecewise Fourier bases for $B(s,t)$, while  Besse \& Cardot (1996) developed spline-based approaches.  Ramsay \& Silverman (1997) presented this model and discussed using basis functions $\bPhi(s)$ and $\bPsi(t)$ for $X_i(s)$ and $Y_i(t)$, respectively, in which case the estimator can be written $B(s,t)=\bPhi(s)' \mathbf{B} \bPsi(t)$, where $\mathbf{B}$ is a $K_x \times K_y$ matrix containing the coefficient surface in the basis space.  They discussed various approaches to regularization, including no regularization, truncation of $X$ and/or $Y$, and bivariate roughness penalties.  All of these methods assumed iid residual errors, no random effect functions, and no modeling of within-function correlation in the errors.

Yao et al. (2005b) presented fPC-based methods for this model where $Y(t)$ and $X(s)$ are modeled using fPC decompositions with iid measurement errors.  The PC scores for $X$ and $Y$ were computed by PACE (Yao et al. 2005a) and estimation of the functional coefficient surface $B(s,t)$ amounts to a series of independent simple linear models regressing estimated response PC scores on predictor PC scores.  Wu \& M\"uller (2011) used WLS to account for the correlation within the functions in estimating the regression coefficients, and Chen et al. (2011) adapted this approach to perform the regression of the sparse, noisy observations $Y_i(t_{ij})$ on the predictor function $X_i(s)$ rather than the estimated PC scores of the denoised estimator.

Ivanescu et al. (2012) extended PFR of Goldsmith et al. (2011a) to the function-on-function setting, using PC expansions of the functional predictors while keeping many PCs, and regularizing the functional coefficient using P-splines.    Their model includes scalar fixed effects, potentially multiple function-on-function regression terms, iid residual errors, and no random effect functions.  Their approach accommodates sparse or densely sampled functions, with or without measurement error.  Scheipl \& Greven (2012) discussed identifiability issues in the context of this model.

Work has been done on two constrained versions of model (\ref{eq:FonF}).  One is the \textit{concurrent model}  (Ramsay \& Silverman 2005), in which  both $X$ and $Y$ are defined on the same domain $t$ and the prediction of $Y_i(t)$ only depends on concurrently observed $X_i(t)$, i.e. $B(s,t)=B(t)$.  This is a special case of the varying coefficient model (Hastie \& Tibshirani 1993).  Wu et al. (1998) described kernel-based methods  with $k$ functional predictors and iid errors, and the functional coefficients smoothed by kernel estimation, and Hoover et al. described smoothing spline and locally weighted polynomial approaches.  Fan \& Zhang (2000) described a two-step approach, first performing pointwise regressions and then smoothing the resulting functional coefficients.  Liang et al. (2003) included measurement errors in the predictor functions using random effect functions, using unstructured covariance.  The functional regression coefficients and random effect functions were both represented by B-spline bases, with penalization done by the truncation inherent in knot selection, and estimated using a WLS-like procedure.  Wu \& Liang (2004) discussed a similar approach that includes random intercepts and slopes for each subject and independent but heteroscedastic errors, and used local polynomial regression to regularize the functional coefficients.  Huang et al. (2004) represented the functional coefficients using B-splines with regularization by the truncation inherent in knot selection, estimating the within function covariance matrix as unstructured in the B-spline basis space.  They accounted for the within-function correlation in inference, but used a least-squares procedure for estimation that ignores this correlation.  Like James et al. (2000), Zhou et al. (2008) represented the mean functions for two longitudinal responses using B-splines regularized by roughness penalties and the respective random effect functions by PC of B-spline coefficients, and then computed the functional surface by regressing the PC scores on each other.

A second constrained model that has been studied is the \textit{historical functional linear model} (hFLM, Malfait \& Ramsay, 2003), for which $B(s,t)$ is lower triangular, allowing only $X_i(s)$ with $s \le t$ to inform prediction of $Y_i(t)$.  Malfait \& Ramsay (2003) represented $B(s,t)$ using custom basis functions $\phi(s,t)$ that are tent-like piecewise linear functions on some grid, constructed to ensure the $s \le t$ constraint, and fit a model with iid errors. Harezlak et al. (2007) considered a variety of regularization techniques for linear B-spline basis functions, including basis truncation, roughness penalties, and sparsity penalties (LASSO). Kim et al. (2011) presented a method using fPC decompositions for both $Y$ and $X$ and assuming iid errors.  The functional coefficient is written out as a product of varying coeffient in $t$ and predetermined basis function such as Fourier, splines, or PCs in $s$, which after integration of $s$ out of the model within the domain $(0,t)$, leads to a varying coefficient model in $t$.  Eigen-decompositions of the marginal variances of $X(s)$ and $Y(t)$ and the cross-covariance of $(X(s),Y(t))$ also come into play in their modeling.

Most work to date has assumed iid errors and no random effect functions, and so does not take within-function or between-function correlations into account in estimation or inference, or incorporate many of the modeling structures developed for functional response regression, with two exceptions.  The functional additive mixed model of Scheipl et al. (2014) discussed in the previous section also accommodates functional predictors, allowing linear regression and two types of additive regression: additive in the PCs like M\"uller \& Yao (2008) or additive in the sense of Hastie \& Tibshirani (1986), effectively generalizing McLean et al. (2012).  Their approach requires discretizing the predictor coefficient on a grid of $s$. %Their construction can accommodate the constraints in the concurrent and historical models when desired. 
 Additionally, Meyer et al. (2013) extended the Bayesian FMM methods of Morris \& Carroll (2006) and following to the function-on-function regression setting, using (potentially different) basis transformations on both the functional responses and predictors and basis truncation, roughness penalties, or sparsity used for regularization.  This model is embedded in the general FMM setting, allowing multiple levels of fixed effect functions for scalar and functional predictors and multiple levels of random effect functions, fit in a way that both within- and between-function variability are accounted for in estimation and inference.  Pointwise and joint functional inference is done on the coefficient surfaces.  %While not applied in their paper, their modeling structure allows nonlinear additive effects $B(x, s, t)$ by adding a level of random effects with Demmler-Reinsch spline design matrix with common variance component for each basis coefficient as in Morris, et al. (2014).

\vspace{-10pt}

\section{Discussion}

There has been a great deal of work on functional regression, especially in the past decade, but clearly there is still much work to be done.  Most of the work to date focuses on 1d continuous-valued functions observed on Euclidean domains, and most methods are designed for smooth functional data on sparse grids as commonly encountered in longitudinal data settings.  The scope of functional data analysis is much broader than this, with many applications yielding 2d or 3d functions such as quantitative image data, functions on non-Euclidean manifolds (e.g. spheres, closed surfaces), discrete- or matrix-values functions, and complex functions with numerous features and sampled on a high dimensional, fine grid.  

A few of the papers highlighted here deal with image data or 2d/3d functions.  Given the emerging importance of these types of data in various applications and the often-times naive analyses commonly used in practice for them, further development in this area is crucial.  There are some preliminary papers involving functional data on manifold domains and a few others dealing with discrete-valued functions, but not much, and so this is another emerging area in the coming years.  Methods that involve general basis function approaches can potentially be directly applied to manifold data in cases where basis functions honoring the inherent constraints can be constructed.  Very few methods have been designed for complex, high-dimensional functional data, which are encountered in a number of emerging applications generating automated measurements on grids of time and/or space.  The Bayesian FMM methods of Morris \& Carroll (2006) and following have been specifically developed with this setting in mind, and has been used for very high dimensional genomic (Morris et al. 2008; Lee \& Morris 2014) and imaging data (Morris et al. 2011, Fazio et al. 2013), in which the functions are complex with many local features and are observed on grids of thousands up to millions.  Most other functional regression methods with available software are at least quadratic in the number of observations per function $T$, so cannot scale up to these enormous sizes.  More functional regression methods need to be developed with these types of complex functions and scalability to enormous functions in mind.

There are many alternative basis functions and regularization approaches used in functional regression analyses.  One open problem is to determine when each type of basis is suitable and preferable.  Principal components are a workhorse for many FDA methods, but more study needs to be done to determine in what settings they should be used.  With complex, high dimensional functions with many features and relatively small sample size, the empirical covariance matrices upon which the PCs are computed are poorly estimated, and it is not clear how many of the eigenfunctions are well estimated, or what their use implies when they are not.  There are preliminary theoretical results on high dimension, low sample size asymptotics (see Section \ref{sec:basis}), but the practical implications of these results need to be explored.  In settings where PCs are not well-estimated, what other bases should be used?  Alternative scalable approaches that can provide some degree of flexibility in representing the covariances but are estimable in such settings would be helpful.  Sparse graphical models may be useful in this context.  Another welcome addition to the literature would be statistically principled approaches to select among a set of candidate bases.  The literature would benefit from a better understanding of regularization and the development of more methods to adaptively regularize data to adapt to local features.  More incorporation of modern sparsity penalties and priors into the functional regression modeling will likely lead to some of these developments.

Statistical principles suggest that regularization in the various ways discussed in this article should lead to more efficient estimation and more accurate inference.  It would be helpful to have careful studies of the importance of regularization of the functional coefficients for estimation and inference in these various models.  How much of a difference does it make?  Is it more important for estimation or inference? Also, it would be helpful to determine in what settings it is more important to account for within-function correlation in estimation of the functional coefficients, how accurately one must represent or estimate the correlation in order to derive a benefit, and for what types of inferential or predictive tasks it is most important.  There are similar relevant questions in terms of accounting for between-function correlation when it is present that seem to be poorly understood by most researchers, and need to be addressed.  These issues are challenging to study theoretically given the vastly different types of functional data encountered in practice (i.e. not all are simple smooth functions), the difficulty to determine the most appropriate asymptotic (e.g. based on number of curves, positions within curve, or numbers of clusters of curves), and the difficulty of assessing whether asymptotic results are relevant to a particular applied setting.   

Many published methods have focused primarily on point estimation.  Although there are approaches in the literature for computing pointwise bands, joint bands, hypothesis tests, and model selection statistics for functional regression, there is still more work that needs to be done in that area, and care needs to be taken so that the inferential methods take into account all of the various within- and between-function sources of variability that characterize these rich data.

\section{Acknowledgements}
This work was supported by NIH grants CA-107304, CA-160736, and CA-016672.

\begin{table}
\centering
{\Large\textbf{Glossary of Abbreviations}}
\begin{tabular}{l l l}
Abbreviation & Definition & First Defined \\
\midrule
AMM & Additive Mixed Models & p. 18 \\
BLUP & Best Linear Unbiased Predictor & p. 4 \\
CAR & Conditional AutoRegressive process & p. 13\\
DWT & Discrete Wavelet Transform & p. 7\\
EM & Expectation Maximization & p. 17\\
FAM & Functional Additive Models & p. 9 \\
FAME & Functional Adaptive Model Estimation & p. 9\\
FANOVA & Functional Analysis of Variance & p. 12\\
FDA & Functional Data Analysis & p. 1\\
FDR & False Discovery Rate & p. 16\\
FGAM & Functional Generalized Additive Model & p. 9\\
FLiRTI& Functional Linear Regression That's Interpretable & p. 7\\
FLM & Functional Linear Model & p. 4\\
FMM & Functional Mixed Model & p. 13\\
fPC(A) & Functional Principal Components (Analysis) & p. 4\\
fPLS & Functional Partial Least Squares & p. 8\\
GAM & Generalized Additive Models & p. 9\\
GCV & Generalized Cross Validation & p. 10\\
GFLM & Generalized Functional Linear Model & p. 5\\
GLASS & Generalized Linear Additive Smooth Structures & p. 7\\
LASSO & Least Absolute Shrinkage and Selection Operator & p. 3\\
LMM & Linear Mixed Models & p. 11\\
MCMC & Markov Chain Monte Carlo & p. 6\\
ML-fPC(A) & Multi-Level functional Principal Components (Analysis) & p 4\\
MPSR & Multidimensional Penalized Signal Regression & p. 7\\
OLS & Ordinary Least Squares & p. 13\\
PACE & Principal Analysis by Conditional Expectation & p. 4\\
PC(A) & Principal Components (Analysis) & p. 3\\
PCR & Principal Component Regression & p. 3\\
PenLS & Penalized Least Squares & p. 3\\
PFR & Penalized Functional Regression & p. 8\\
PLS & Partial Least Squares & p. 6\\
PSR & Penalized Signal Regression & p. 6\\
RJMCMC & Reversible Jump Markov Chain Monte Carlo & p. 11\\
RLRT & Restricted Likelihood Ratio Test & p. 6\\
SCAD & Smoothly Clipped Absolute Deviation & p. 7\\
SPSR & Space-Varying Penalized Signal Regression & p. 7\\
SSVS & Stochastic Search Variable Selection & p. 3\\
WLS & Weighted Least Squares & p. 10
% add CCA, DTI, FA, MALDI-TOF, MPS, MS, RCST if needed
\end{tabular}
\end{table}
\newpage
\bibliographystyle{unsrt}

\end{document}